\documentclass[prc,twocolumn]{revtex4}
\usepackage{graphicx}
\newcommand{\D}{\displaystyle}

\makeatletter
  \newcommand{\figcaption}{\def\@captype{figure}\caption}
  \newcommand{\tabcaption}{\def\@captype{table}\caption}
\makeatother

\begin{document}
\title {A systematic study of neutron magic nuclei with $N$ = 8, 20, 28, 50, 82, and 126 \\
in the relativistic mean field theory}

\author{
L. S. Geng,$^{1,2}$ 
  H. Toki,$^{1}$ 
  and J. Meng$^{2}$
}

\affiliation{
$^1$Research Center for Nuclear Physics (RCNP), Osaka University,
Ibaraki, Osaka 567-0047, Japan
\\
$^2$School of Physics, Peking University, Beijing 100871, People's
Republic of China }

\begin{abstract}
We perform a systematic study of all the traditional neutron magic
nuclei with $N$ = 8, 20, 28, 50, 82, and 126, from the neutron
drip line to the proton drip line. We adopt the deformed
relativistic mean field (RMF) theory as our framework and treat
pairing correlations by a simple BCS method with a zero-range
$\delta$-force. Remarkable agreement with the available
experimental data is obtained for the binding energies, the two-
and one-proton separation energies, and the nuclear charge radii.
The calculated nuclear deformations are compared with the
available experimental data and the predictions of the FRDM mass
formula and the HFBCS-1 mass formula. We discuss, in particular,
the appearance of sub-shell magic nuclei by observing irregular
behavior in the two- and one-proton separation energies.
\end{abstract}

\maketitle

\section{Introduction}

``Magic number" is a very important concept in many subjects of
physics, such as atomic physics, nuclear physics and micro cluster
physics. In nuclear physics, nuclei with magic numbers have been
hot topics of nuclear research since the beginning of this subject
\cite{mayer.49,jensen.49}. In recent experiments with radioactive
nuclear beams (RNB), disappearance of traditional magic numbers
and appearance of new magic numbers are observed in nuclei with
exotic isospin ratios. Furthermore, nuclei with magic neutron
numbers in the neutron-rich region are of special interest for the
study of astrophysical r-process \cite{burbidge.57}.

The unusual stability of nuclei with neutron (proton) numbers 2,
8, 20, 28, 50, 82, and 126, commonly referred to as ``magic
numbers'', was traditionally explained in the nonrelativistic
shell model approximated by the 3D Harmonic-Oscillator central
potential together with a very strong spin-orbit interaction
introduced by hand
\cite{mayer.49,jensen.49,bohr.69,heyde.99,ring.80}. On the other
hand, in the models within the relativistic framework, say the
relativistic mean field (RMF) theory
\cite{walecka.74,serot.86,reinhard.89,toki.91,ring.96}, the strong
spin-orbit interaction appears naturally as the interplay between
the strong scalar and vector potentials, which are necessary for
reproducing the saturation properties of nuclear matter. Due to
the proper setting of the scalar and the vector potentials the
shell structure is obtained without any additional parameters for
the spin-orbit splittings.

Recently, it is argued that the magic numbers are of a localized
feature, i.e. in nuclei with exotic isospin ratios the classical
magic numbers do not necessarily hold and new magic numbers may
appear. There have been several experimental evidences supporting
such a belief, including the lately reported $N=82$ shell
quenching \cite{dillmann.03} and the appearance of a new neutron
magic number $N=16$ in the neutron-rich light nuclei
\cite{ozawa.00}.  Such a localized feature of magic numbers are
also claimed to be important for various nuclear-astrophysical
problems \cite{kratz.93}. The disappearance or quenching of
nuclear magicity near both neutron and proton drip lines have been
discussed quite a lot in various nuclear models, including the
infinite nuclear matter (INM) model \cite{nayak.99}, the extended
Bethe-Weizs\"{a}cker mass formula \cite{samanta.02}, the
antisymmetrized molecular dynamics (AMD) \cite{horiuchi.02}, the
Hartree-Fock method \cite{chen.95}, and the relativistic mean
field (RMF) \cite{MTY98,geng.033} theory.  In the relativistic
mean field model, while the $Z =$ 8, 20, 28, 50, 82, and 126
isotopic chains have been discussed a lot, a systematic study of
all the nuclei with neutron magic numbers is still missing.

The appearance and disappearance of the magic number effect are
intimately related with the spin-orbit interaction and the onset
of deformation.  In this sense, a systematic study of nuclei in
various mass regions in terms of the relativistic mean field
theory is suited, since the spin-orbit interaction arises
naturally in relation with the saturation property.  Hence, we are
able to provide the variation of the spin-orbit splittings with
the proton and neutron numbers.  As for the deformation, nuclei
with magic numbers are not easy to acquire deformation, because
the onset of deformation does not utilize the energy gain coming
from the large shell gaps.  At the same time, due to this reason
it is easier to find the appearance of the sub-magic effects for
nuclei with either the proton number or the neutron number being
magic numbers.

In the present work, we would like to make a systematic study of
neutron magic nuclei from the neutron drip line to the proton drip
line in terms of the deformed relativistic mean field theory.  The
pairing correlations are also very important to make the nucleus
tend to be spherical. Therefore, we take the recently proposed
method of using the zero-range $\delta$-force for the pairing
correlations in order to pick up the resonant states in the
continuum when the nucleus approaches the neutron threshold
\cite{geng.03}. The quadrupole constrained RMF calculation is
performed for each nucleus to find all the energy minima as a
function of the nuclear deformation. In this paper, we stay our
calculations at the mean-field level and hence anticipate that we
may miss the onset of small deformation in this framework for the
case, in which the deformation effect is not dominant. Hence, the
purpose of this paper is to make a systematic study of the neutron
magic nuclei in the mean field framework and compare with the
available experimental data in the global sense and with other
more sophisticated theoretical models for particular nuclei. We
mention that even at the mean-field level, many proton magic
nuclei appear to be deformed \cite{geng.033}.

This paper is organized as follows. We provide a short summary of
the deformed RMF+BCS method in Sec. II. In Sec. III-VI, we present
the results of our calculations, including the binding energies,
the two- and one-proton separation energies, the nuclear radii,
and the nuclear deformations. With these quantities, we also
discuss the magicity of these neutron magic nuclei with $N=8$, 20,
28, 50, 82, and 126. The whole work is summarized in Sec. VII.

\section{Theoretical framework}

In the present work, the recently proposed deformed RMF+BCS method
with a zero-range $\delta$-force in the pairing channel is adopted
\cite{geng.03}. The zero-range $\delta$-force has proved to be
very successful to take into account the continuum effect by
picking up the resonant states both in relativistic and
nonrelativistic self-consistent mean field models
\cite{yadav.02,sand.03,geng.03,meng.98,meng98prl,sand.00}. In the
mean-field part, the TMA parameter set is used \cite{sugahara.94}.
The RMF calculations have been carried out using the model
Lagrangian density with nonlinear terms for both $\sigma$ and
$\omega$ mesons as described in detail in Refs.
\cite{geng.03,sugahara.94}, which is given by
\begin{eqnarray}
\mathcal{L} &=& \bar \psi (i\gamma^\mu\partial_\mu -M) \psi \nonumber\\
&+&\,\frac{\D 1}{\D
2}\partial_\mu\sigma\partial^\mu\sigma-\frac{\D 1}{\D
2}m_{\sigma}^{2} \sigma^{2}- \frac{\D 1}{ \D
3}g_{2}\sigma^{3}-\frac{\D 1}{\D
4}g_{3}\sigma^{4}-g_{\sigma}\bar\psi
\sigma \psi\nonumber\\
&-&\frac{\D 1}{\D 4}\Omega_{\mu\nu}\Omega^{\mu\nu}+\frac{\D 1}{\D
2}m_\omega^2\omega_\mu\omega^\mu +\frac{\D 1}{\D
4}g_4(\omega_\mu\omega^\mu)^2-g_{\omega}\bar\psi
\gamma^\mu \psi\omega_\mu\nonumber\\
 &-& \frac{\D 1}{\D 4}{R^a}_{\mu\nu}{R^a}^{\mu\nu} +
 \frac{\D 1}{\D 2}m_{\rho}^{2}
 \rho^a_{\mu}\rho^{a\mu}
     -g_{\rho}\bar\psi\gamma_\mu\tau^a \psi\rho^{\mu a} \nonumber\\
      &-& \frac{\D 1}{\D 4}F_{\mu\nu}F^{\mu\nu} -e \bar\psi
      \gamma_\mu\frac{\D 1-\tau_3}{\D 2}A^\mu
      \psi,
\end{eqnarray}
where all symbols have their usual meanings. The corresponding
Dirac equations for nucleons and Klein-Gordon equations for mesons
obtained with the mean-field approximation are solved by the
expansion method on the widely used axially deformed
Harmonic-Oscillator basis \cite{geng.03,gambhir.90}. The number of
shells used for expansion is chosen as $N_f=N_b=20$. More shells
have been tested for convergence considerations. The quadrupole
constrained calculations have been performed for all the nuclei
considered here in order to obtain their potential energy surfaces
(PESs) and determine the corresponding ground-state deformations
\cite{geng.03,flocard.73}. For nuclei with odd number of nucleons,
a simple blocking method without breaking the time-reversal
symmetry is adopted \cite{geng.032,ring.80}. The pairing strength
$V_0$ is taken to be the same for both protons and neutrons, but
optimized for different regions by fitting the experimental two-
and one-proton separation energies, more specifically, for the
$N=8$, 20, 28, 50 isotonic chains, $V_0=344.1$ MeV fm$^3$; for the
$N=82$, 126 isotonic chains, $V_0=310$ MeV fm$^3$.

Whenever the zero-range $\delta$ force is used either in the BCS
or the Bogoliubov framework, a cutoff procedure must be applied,
i.e. the space of the single-particle states where the pairing
interaction is active must be truncated. This is not only to
simplify the numerical calculation but also to simulate the
finite-range (more precisely, long-range) nature of the pairing
interaction in a phenomenological way \cite{doba.96,hfb2.02}. In
the present work, the single-particle states subject to the
pairing interaction are confined to the region satisfying
\begin{equation}
\epsilon_i-\lambda\le E_\mathrm{cut},
 \end{equation}
 where $\epsilon_i$ is the single-particle energy, $\lambda$
 the Fermi energy, and $E_\mathrm{cut}=8.0$ MeV. We find that increasing
 $E_\mathrm{cut}$ from 8.0 MeV up to 16.0 MeV, followed by a
 readjustment of the pairing strength $V_0$, does not change the
 results appreciably, and therefore none of our conclusions
 will change. Recently, such a cutoff issue has been discussed in much
 detail by Goriely et al. in the Hartree-Fock-Bogoliubov (HFB) framework \cite{hfb2.02}. Our
cutoff procedure happens to coincide with their conclusions (see
Table II of Ref. \cite{hfb2.02}),
 although a bit different from their final optimal choice used in the construction of the HFB-2 mass table.

 Even after the pairing interaction between like nucleons is taken into account,
 mean-field calculations (HF or RMF) still miss some residual correlations, one of which is the so-called Wigner effect. Such an
 effect is known to be important for light $N\approx Z$ nuclei  by
 making them more bound by about 2 MeV. Although it is usually approximated in
 a phenomenological way in most mass tables
 \cite{moller.95,hfbcs1.01,hfb1.02,hfb2.02,hfb3.03}, here we will ignore this effect due
 to the following considerations. It is expected that these
 residual correlations (including the Wigner effect) have only
 small contributions and are active only under special situations. In the case of
 the Wigner effect, it becomes appreciable only for light nuclei
 with $N\approx Z$.

Finally, the center-of-mass correction is approximated by
 \begin{equation}
 E_{\textrm{cm}}=-\frac{3}{4}41A^{-1/3},
 \end{equation}
which is often used in the relativistic mean field theory among
the many recipes for the center-of-mass correction
\cite{bender.00}. Since adopting new schemes implies that one has
to readjust the model parameters, we will adhere to this simple
approximation in the present work.

\section{Binding energies}
\begin{figure*}[htpb!] \centering
\includegraphics[scale=0.23]{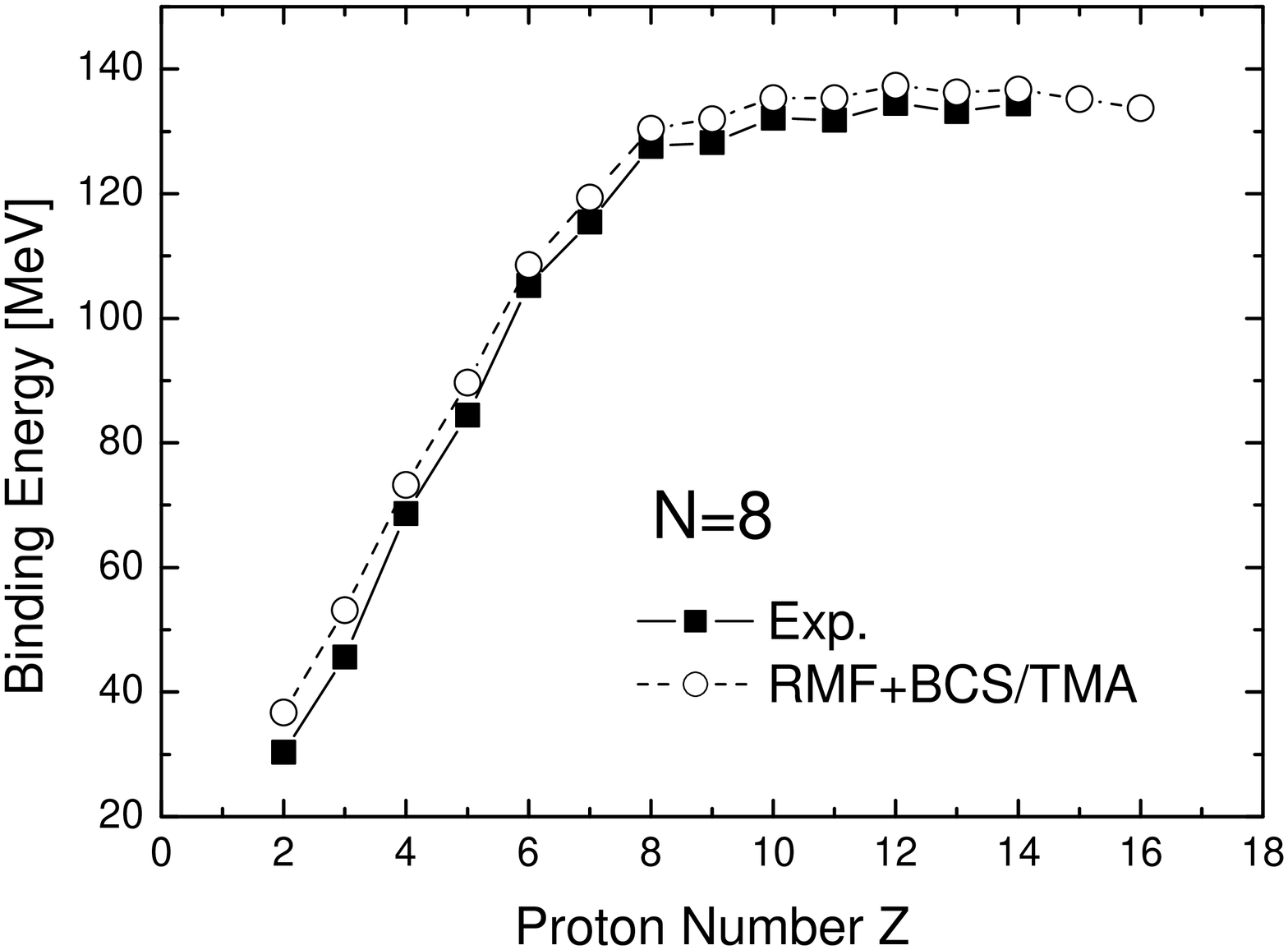}%
\includegraphics[scale=0.23]{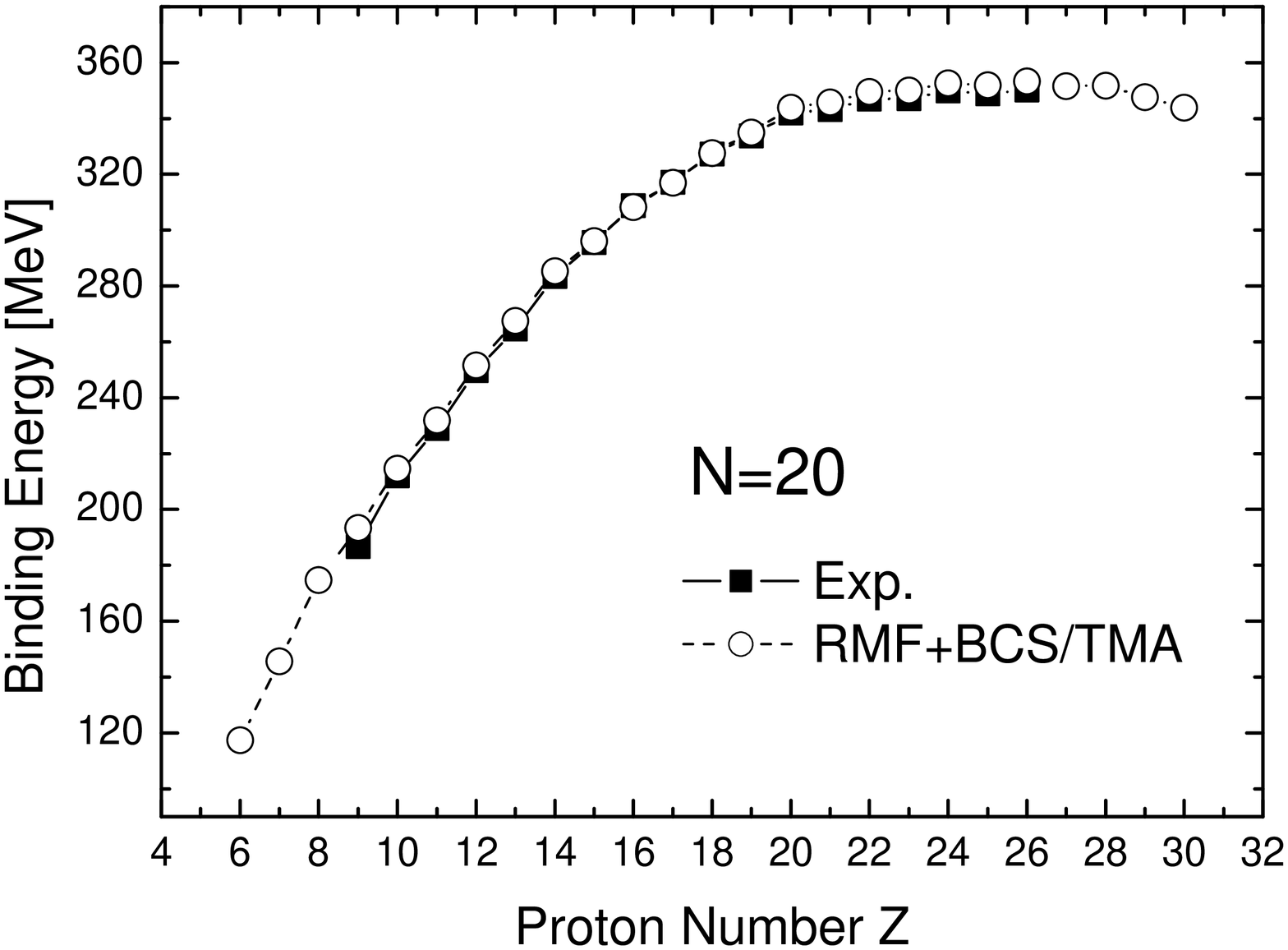}%
\includegraphics[scale=0.23]{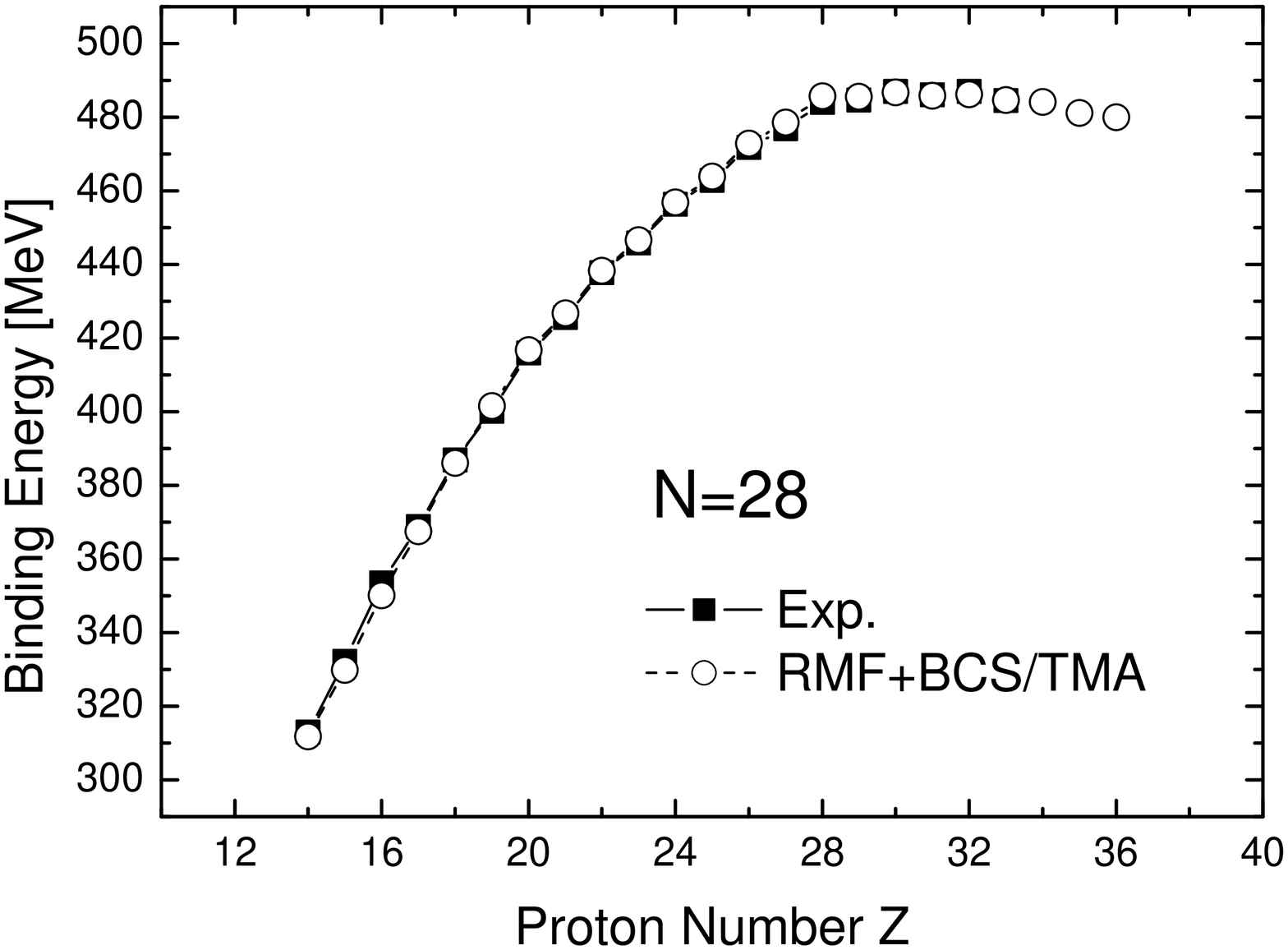}
\includegraphics[scale=0.23]{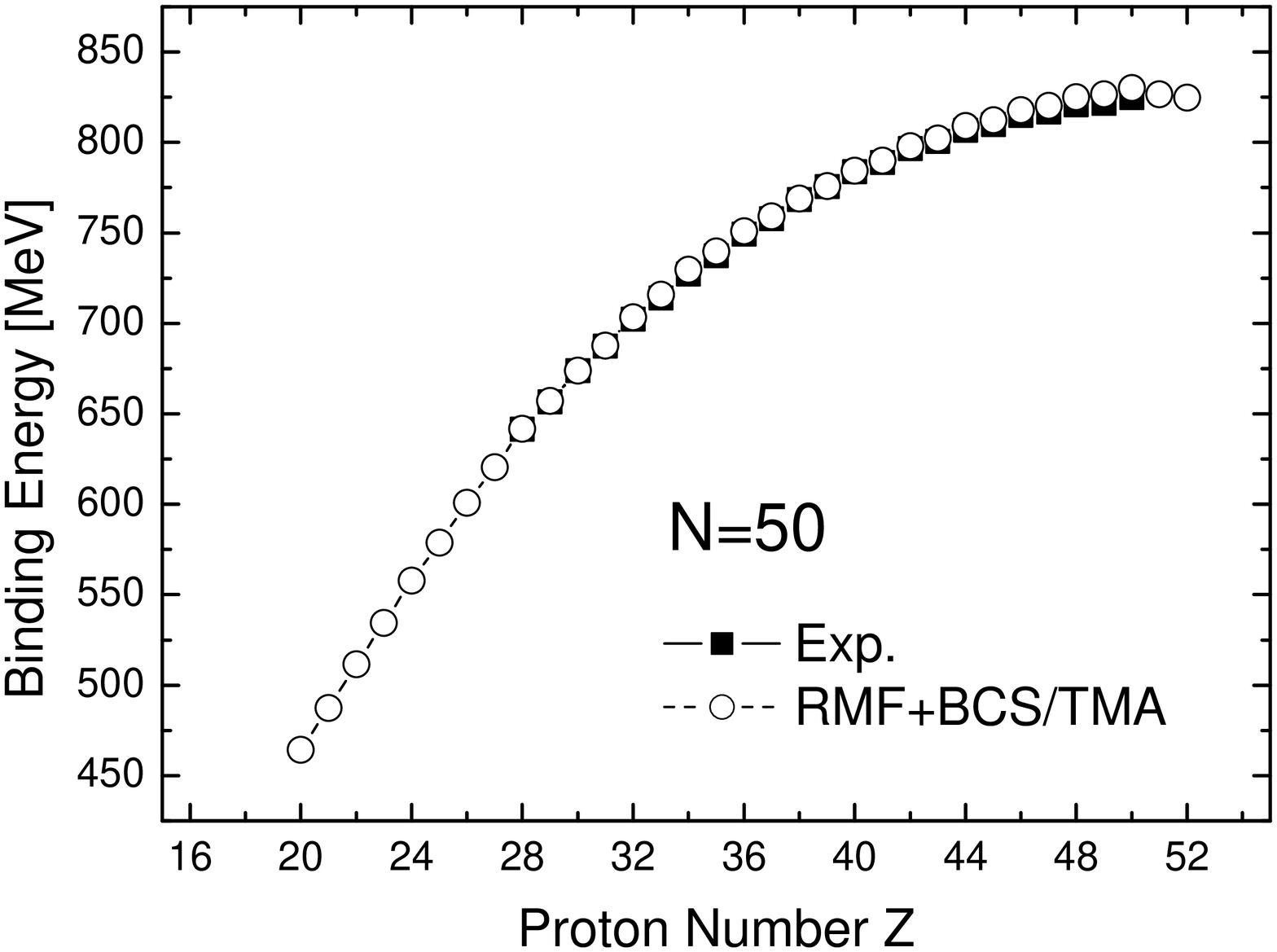}%
\includegraphics[scale=0.23]{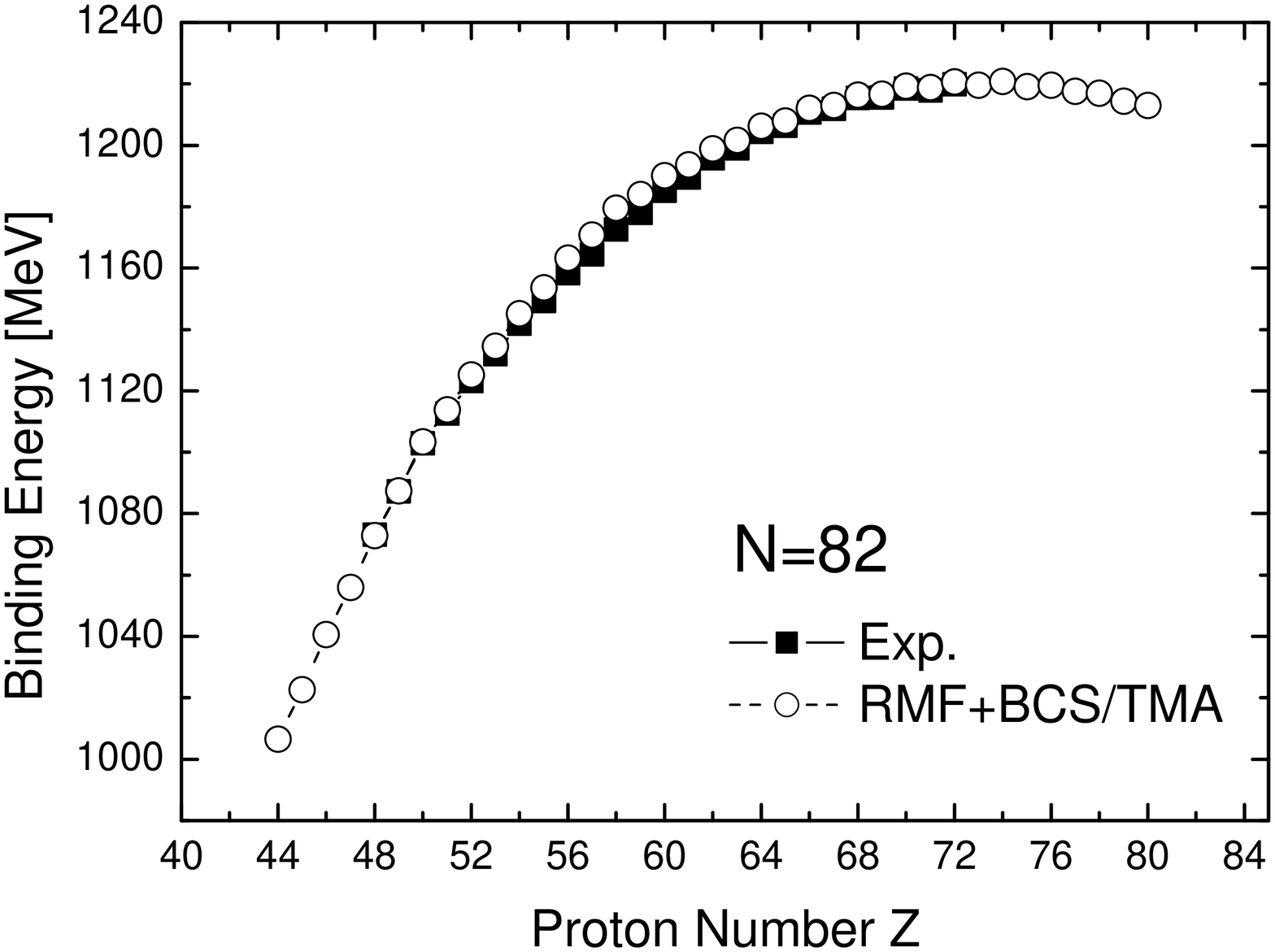}%
\includegraphics[scale=0.23]{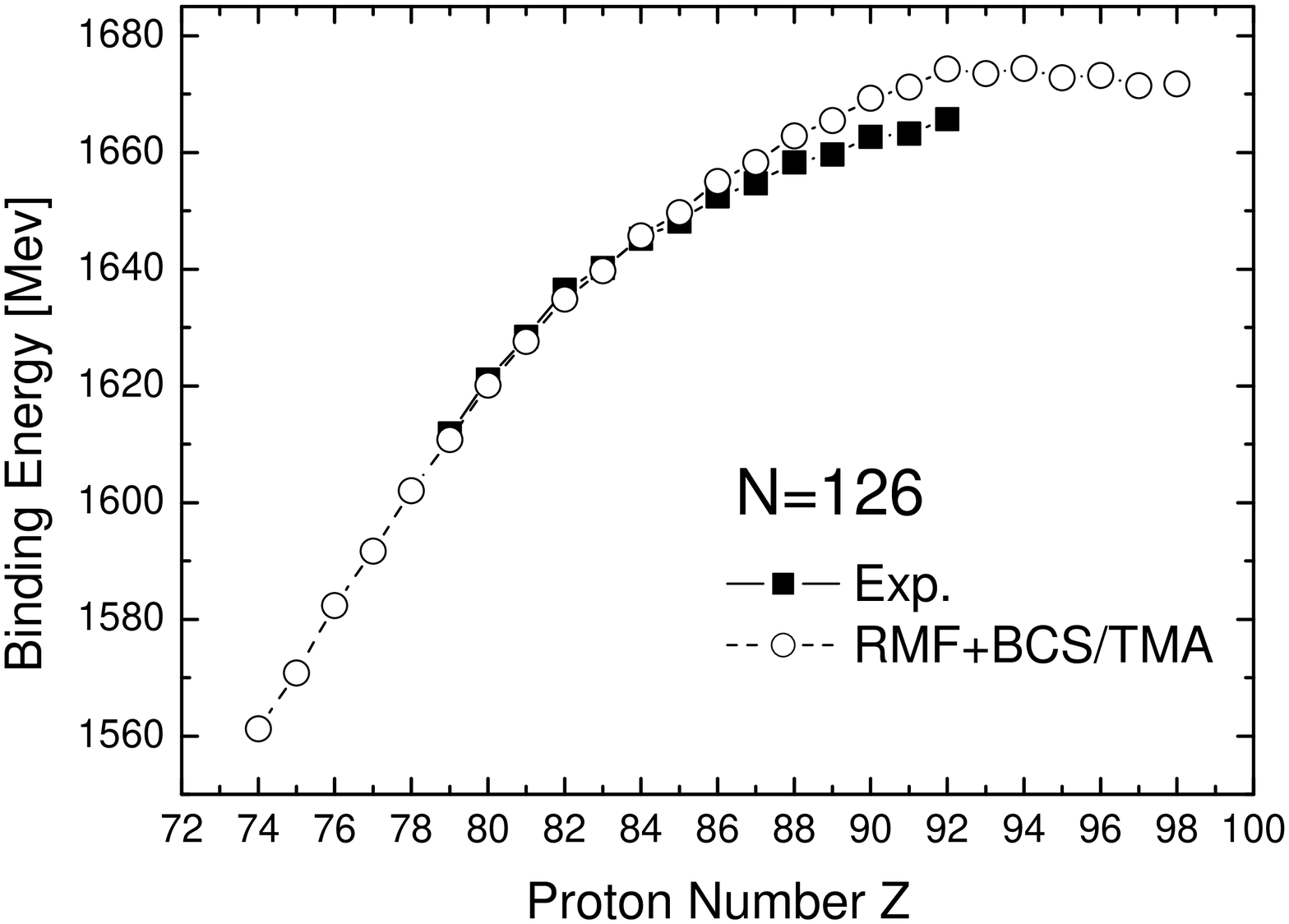}
\caption{The binding energies of the $N=8$, 20, 28, 50, 82, and
126 isotones as functions of the proton number, $Z$. The results
obtained from the deformed RMF+BCS calculations with the TMA
parameter set (open circle) are compared with the available
experimental data (solid square) \cite{audi.95}.}
\end{figure*}

The nuclear binding energy is one of the most basic properties of
nuclei. Although the two- and one-proton separation energies are
more useful for the purpose of identifying the shell structure,
the binding energies of nuclei can also show some important
features of a theoretical model, such as the reliability of the
extrapolation to the unknown areas. In Fig. 1, we compare the
binding energies of all the isotones with neutron number $N=8$,
20, 28, 50, 82, and 126 with the available experimental data
\cite{audi.95}. Remarkable agreement between theory and experiment
can be clearly seen. For the $N=8$ isotonic chain, the
experimental trend is reproduced quite well. The only small
discrepancy is that the calculated results are somewhat
systematically larger than the experimental data. Nevertheless, it
is surprising that a mean field model can reproduce such light
nuclei so well without fitting the parameters particularly for
this region. For the $N=126$ isotonic chain, theory agrees well
with experiment around $^{208}$Pb. The results for nuclei with
more protons begin to deviate from the experimental data, i.e. the
calculated results are larger than the experimental data. The use
of other parameter sets, NL3 \cite{lalazissis.97} and NL-Z2
\cite{bender.99}, does not change this conclusion.

It is well known that the modern HF or RMF calculations still can
not provide a description of nuclear masses comparable to those of
most mass tables \cite{patyk.99}. This feature is attributed to
the limited number of nuclear masses taken into account in the
fitting procedure of the model parameters. On the other hand, when
more masses are included into the fitting procedure as what has
been done by Goriely et al.
\cite{hfbcs1.01,hfb1.02,hfb2.02,hfb3.03}, a description of nuclear
masses comparable to that of the finite-range droplet model (FRDM)
\cite{moller.95} has been obtained based on the HF
\cite{hfbcs1.01} and HFB \cite{hfb1.02,hfb2.02,hfb3.03} methods.
In Table I, we compare the root-mean-square (rms) deviation
$\sigma$ of our present calculation, with that of the HFBCS-1 mass
formula, and that of the FRDM model for 107 nuclei whose
experimental (or extrapolated) masses are compiled in Ref.
\cite{audi.95}.
\begin{table}[htbp]
\setlength{\tabcolsep}{1 em}\caption{The mass rms deviation
$\sigma$  for 107 nuclei with $N=8,20,28,50,82,$ and $126$ from
the present calculation (RMF+BCS), the HFBCS-1 mass formula
(HFBCS-1) \cite{hfbcs1.01}, and the FRDM mass formula (FRDM)
\cite{moller.95}. }
\begin{center}
\begin{tabular}{cccc}
\hline\hline
&RMF+BCS&HFBCS-1\cite{hfbcs1.01}&FRDM\cite{moller.95}\\
\hline
 $\sigma$&2.873&0.957&0.774\\
 \hline\hline
\end{tabular}
\end{center}
\end{table}

It is easily seen that the rms deviation of the present RMF
calculation is still 3 times larger than that of the HFBCS-1 mass
formula or 4 times larger than that of the FRDM model. It should
be noted that the rms deviations for the HFBCS-1 mass formula and
for the FRDM mass formula shown in Table I are larger than their
overall deviations, 0.738 MeV and 0.669 MeV, respectively, mainly
due to the relatively large discrepancies for those light isotones
with $N=8,20$, and $28$. It is expected that a better description
of nuclear masses in the RMF framework be obtained if more masses
are taken into account in the parameter fitting procedure.

\begin{figure*}[t]
\centering
\includegraphics[scale=0.23]{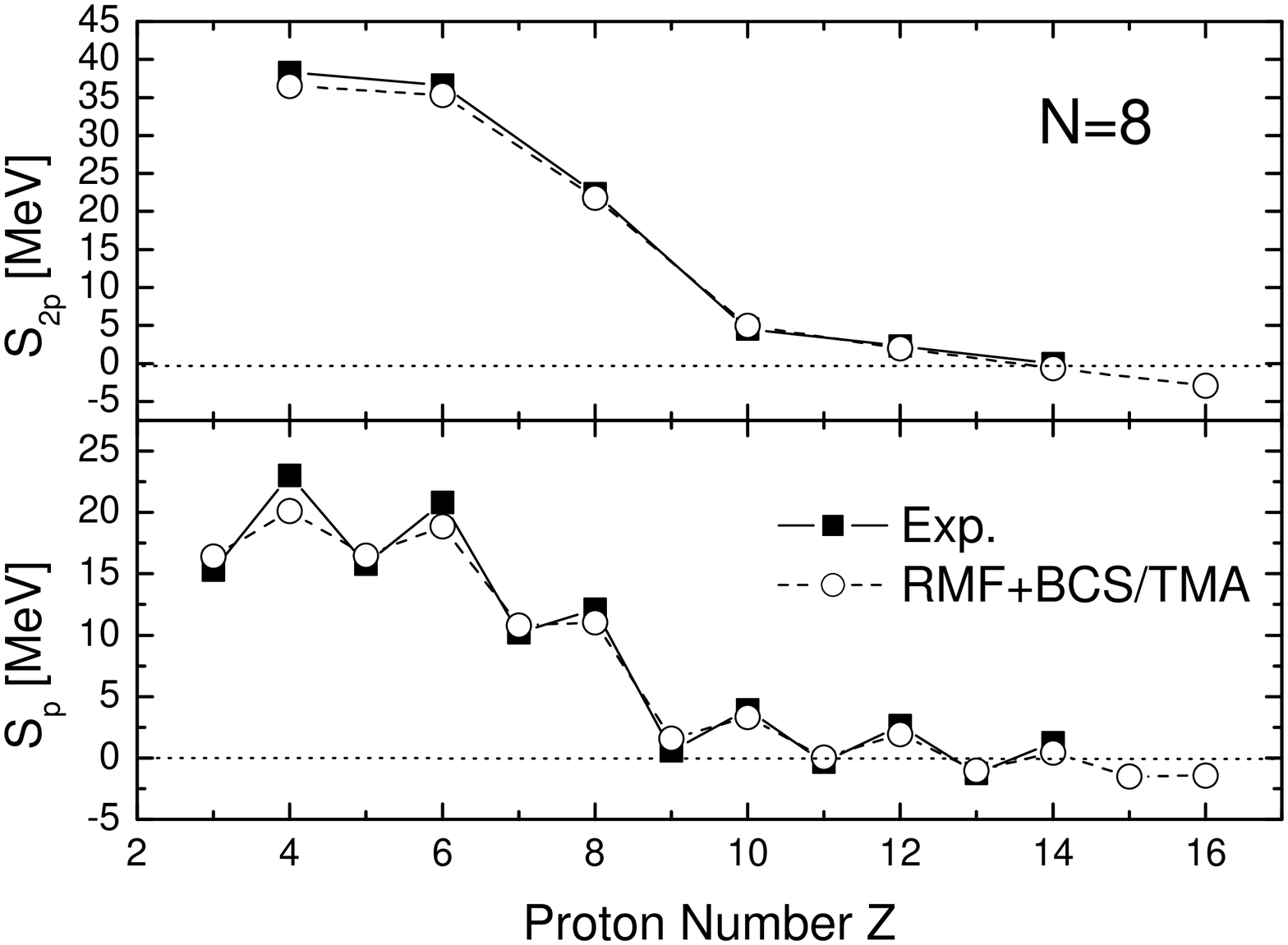}%
\includegraphics[scale=0.23]{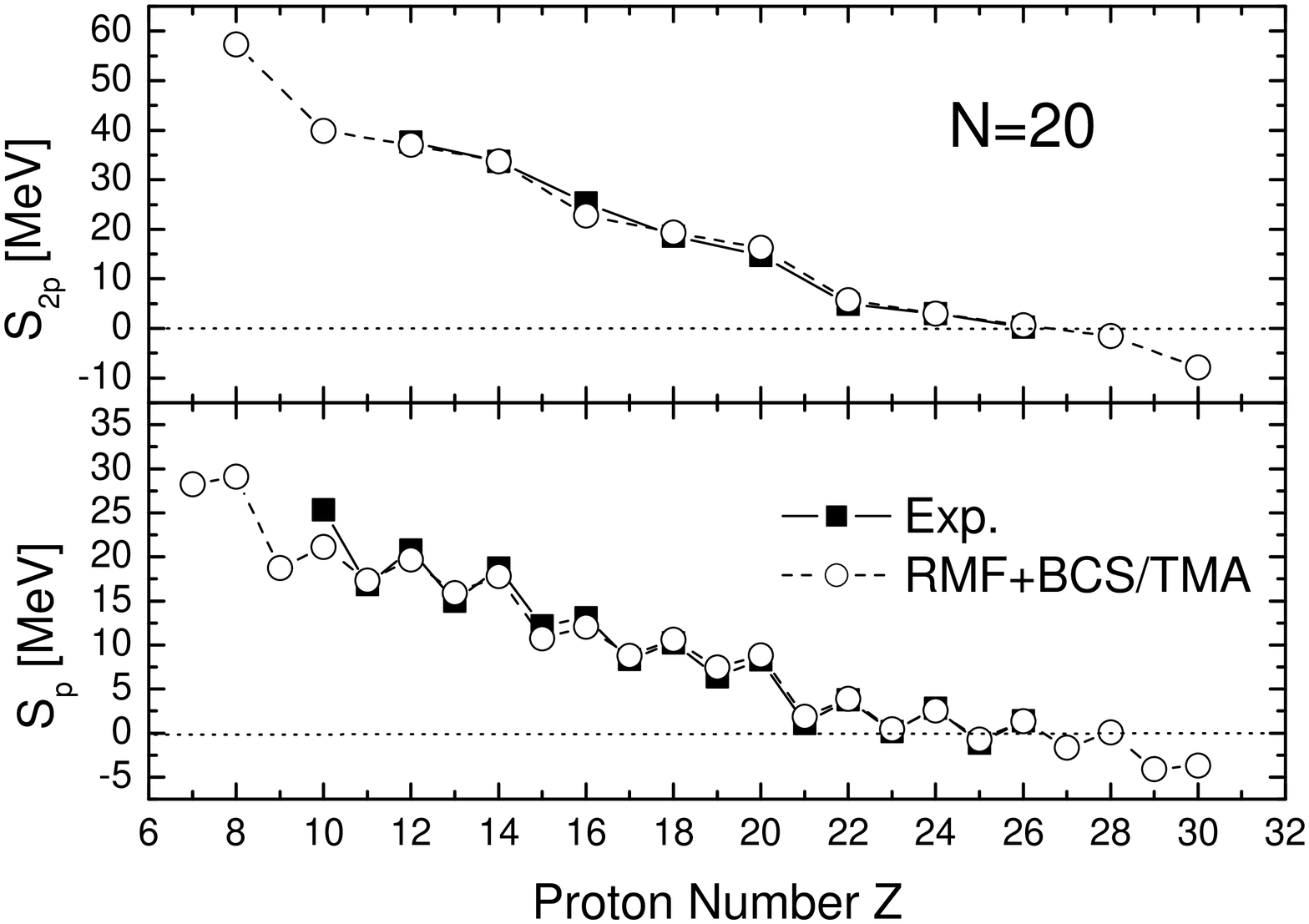}%
\includegraphics[scale=0.23]{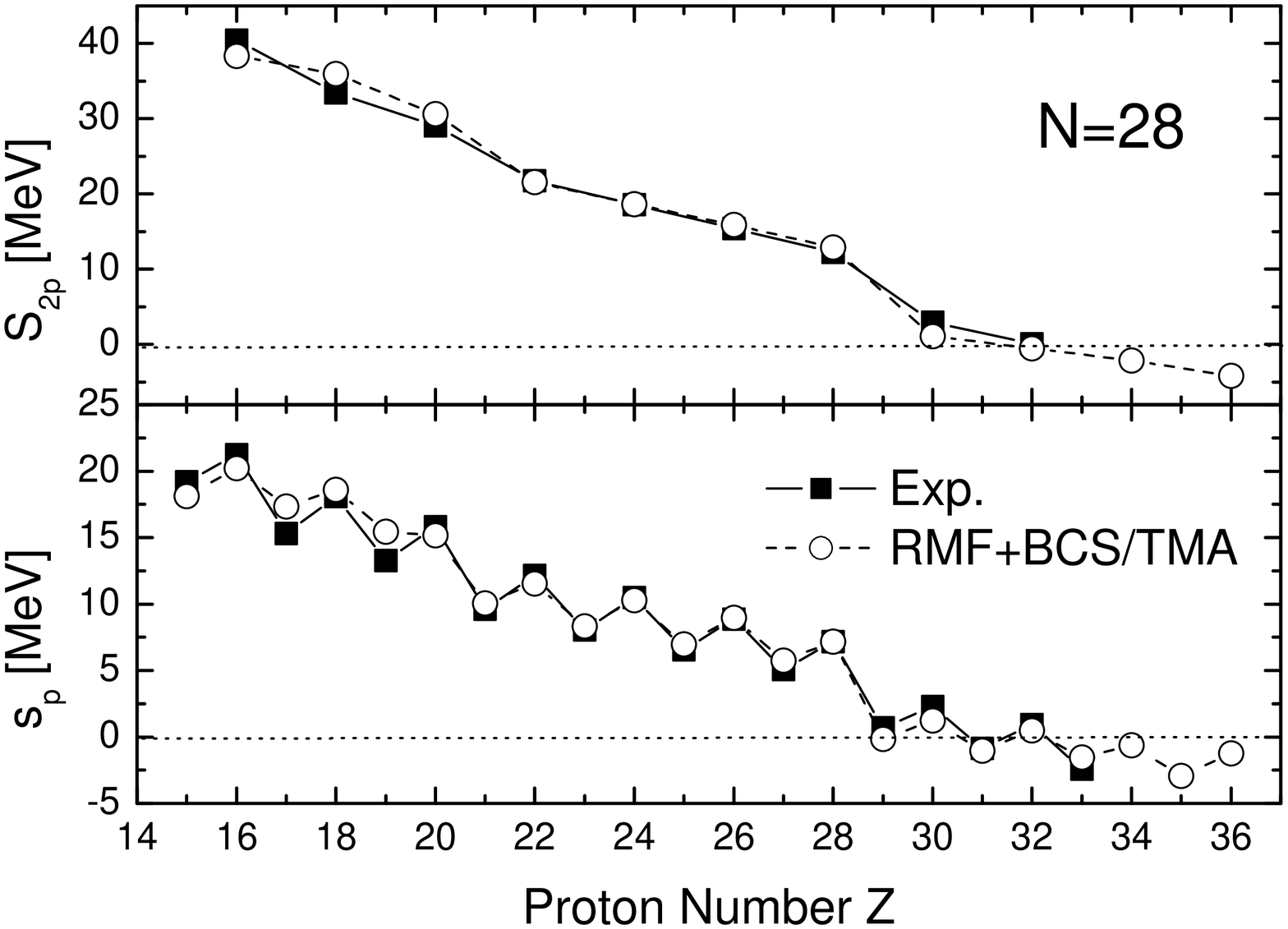}
\includegraphics[scale=0.23]{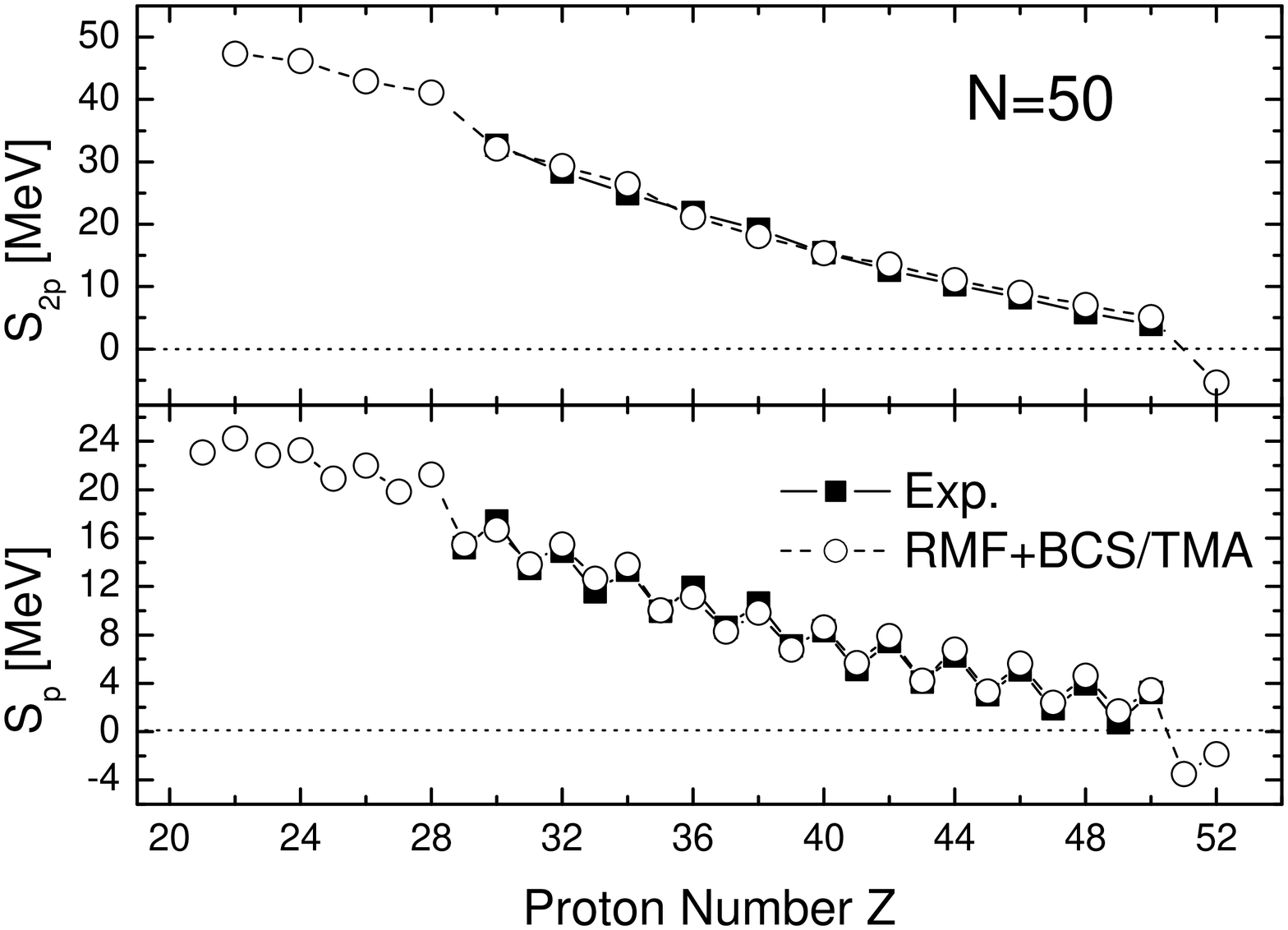}%
\includegraphics[scale=0.23]{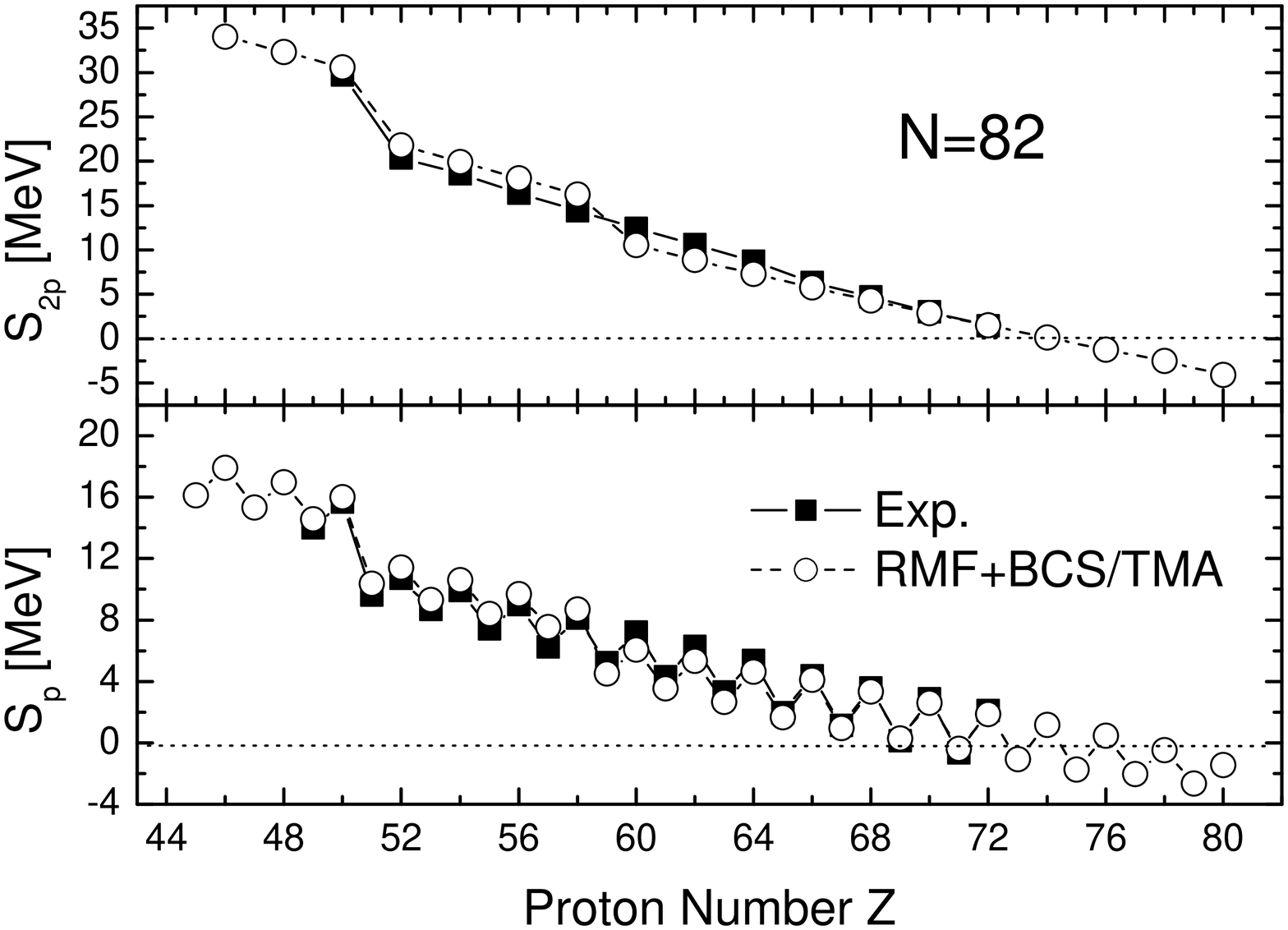}%
\includegraphics[scale=0.23]{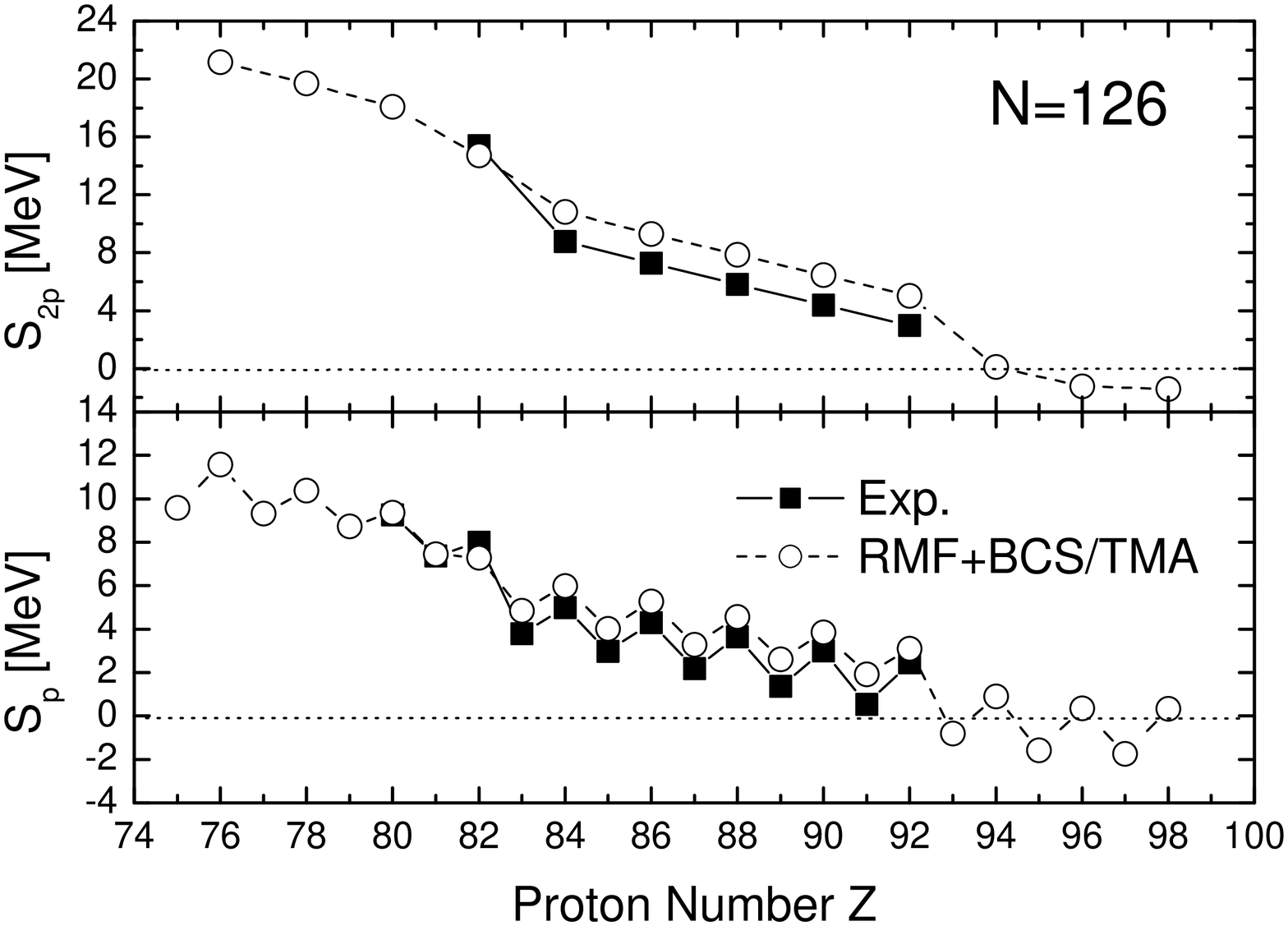}
\caption{The two- and one-proton separation energies, $S_{2p}$ and
$S_p$, of the $N=8$, 20, 28, 50, 82, and 126 isotones as functions
of the proton number, $Z$. The results obtained from the deformed
RMF+BCS calculations with the TMA parameter set (open circle) are
compared with the available experimental data (solid square)
\cite{audi.95}.}
\end{figure*}

\section{Two- and one-proton separation energies}

The two- and one-proton separation energies
 \begin{equation}
 S_{2p}(N,Z)=B(N,Z)-B(N,Z-2),
 \end{equation}
  \begin{equation}
 S_{p}(N,Z)=B(N,Z)-B(N,Z-1),
 \end{equation}
where $B(N,Z)$ is the binding energy of the nucleus with neutron
number $N$ and proton number $Z$, are quite important and
sensitive quantities to describe nuclei and the corresponding
shell closures. In Fig. 2, we plot the two- and one-proton
separation energies for all the $N=8$, 20, 28, 50, 82, and $126$
isotones as functions of the proton number $Z$. A larger drop than
its neighboring counterparts in these curves is usually
interpreted as a (sub)shell closure, and the corresponding nucleon
number is called ``(sub-)magic number''. In the present work, we
mainly adopt such a notion to discuss shell closures.

In Fig. 2, the calculated two- and one-proton separation energies
are compared with the available experimental data \cite{audi.95}.
We should mention that some experimental data from Ref.
\cite{audi.95} are obtained by ``distant connection'' and that we
do not make any distinction between real experimental data and
extrapolated data. From Fig. 2, it is quite clear that the
theoretical calculations agree remarkably well with the
experimental data in the whole region from mass number 11
($^{11}_{3}$Li$_8$) to mass number 218 ($^{218}_{92}$U$_{126}$).

For the $N=8$ isotonic chain, the $Z=8$ ($1p_{1/2}$) shell closure
is clearly seen from both the two- and the one-neutron separation
energies, where the shell quantum number in the bracket is that of
the shell closure. Both experimental data and our calculations
also show a distinct shell closure at $Z=6$ ($1p_{3/2}$). For the
$N=20$ isotonic chain, in addition to the traditional $Z=8$
($1p_{1/2}$), $Z=20$ ($1d_{3/2}$), and $Z=28$ ($1f_{7/2}$) shell
closures, a new (sub)shell closure ($1d_{5/2}$) is clearly seen at
$Z=14$ . $^{34}_{14}$Si$_{20}$ has been recognized as a doubly
magic nucleus by a previous shell-model calculation
\cite{caurier.98}, due to a feature characteristic of doubly magic
nuclei, i.e. it has a $0p-0h$ ground state and its two lowest
excited states are intruders. For the $N=28$ isotonic chain, two
neutron shell closures at $Z=20$ ($1d_{3/2}$) and $Z=28$
($1f_{7/2}$) are reproduced quite well. For the $N=50$ isotonic
chain, the agreement with the available experimental data is very
good. Although the $Z=28$ ($1f_{7/2}$) and $Z=50$ ($1g_{9/2}$)
shell closures are still unaccessible experimentally up to now,
they are clearly shown in our calculations. Based on the overall
good agreement of the $N=50$ isotonic chain, we believe that our
predictions are reliable.

For the $N=82$ isotonic chain, the $Z=50$ ($1g_{9/2}$) shell
closure is reproduced quite well while the $Z=82$ ($1h_{11/2}$)
shell closure is already in the unbound region. We note that there
is a new (sub)shell closure at $Z=58$ ($1g_{7/2}$), as shown in
Fig. 2. It was argued long time ago that there is a whole
``plateau'' of stability for all the even $58\le Z\le 70$ nuclei
in the $N=82$ isotonic chain \cite{abbas.84, arumugam.03}, i.e.
the so-called ``changing magicities''. Our calculations do
indicate this (sub)shell closure, which is also supported by the
experimental data, as shown in Fig. 2. However, the different
deviation trends for nuclei with $Z>58$ and $Z<58$ indicate that
some important features around $Z=58$ in the $N=82$ isotonic chain
may be missed and/or mistreated in the relativistic mean field
theory.

For the $N=126$ isotonic chain, immediately, we notice the
seemingly big deviations of our calculations from the experimental
data. From both the two- and the one-proton separation energies,
we note that the experimental $Z=82$ ($1h_{11/2}$) shell closure
is underestimated somewhat by the relativistic mean field
calculations. Our calculations also indicate another new
(sub)shell closure ($1h_{9/2}$) at $Z=92$ in agreement with Ref.
\cite{rutz.98}, where $^{218}_{92}$U$_{126}$ is studied as a
doubly magic nucleus. We notice that this is a unique phenomenon
in the relativistic mean field theory while a large-scale
shell-model calculation did not find this shell closure
\cite{caurier.03}. We also find that the use of other often used
parameter sets in this region, such as NL3 \cite{lalazissis.97}
and NL-Z2 \cite{bender.99}, does not change our conclusion.

The two-proton separation energy becomes negative when the nucleus
becomes unstable with respect to two-proton  emission. Hence, the
two-proton drip-line nucleus for the corresponding isotonic chain
is the one with two less protons than the nucleus at which
$S_{2p}$ first becomes negative. In the same way, we can also name
the one-proton drip-line nucleus. The predicted two- and
one-proton drip-line nuclei based on our calculations are
$^{20}_{12}$Mg$_{8}$; $^{46}_{26}$Fe$_{20}$,
$^{44}_{24}$Cr$_{20}$; $^{58}_{30}$Zn$_{28}$,
$^{56}_{28}$Ni$_{28}$; $^{100}_{50}$Sn$_{50}$;
$^{156}_{74}$W$_{82}$, $^{152}_{70}$Yb$_{82}$;
$^{220}_{94}$Pu$_{126}$; $^{218}_{92}$U$_{126}$; respectively. It
is interesting to note that in some cases the two-proton drip-line
nucleus and the one-proton drip-line nucleus are the same one,
such as $^{20}_{12}$Mg$_{8}$ in the $N=8$ isotonic chain and
$^{100}_{50}$Sn$_{50}$ in the $N=50$ isotonic chain. In other
cases, the one-proton drip-line nucleus comes before the
two-proton drip-line nucleus, which is easy to understand because
the pairing interaction can increase the stability of even-even
nuclei compared with its one-neutron less isotone. Depending on
the pairing strength of the specific situation, the difference in
neutron number between the two- and one-proton drip-line nucleus
could be two or four. More specifically, the difference is 2 in
the $N=20$, 28 and 126 isotonic chains, while it is 4 in the
$N=82$ isotonic chain.

\begin{figure*}[htpb!] \centering
\includegraphics[scale=0.23]{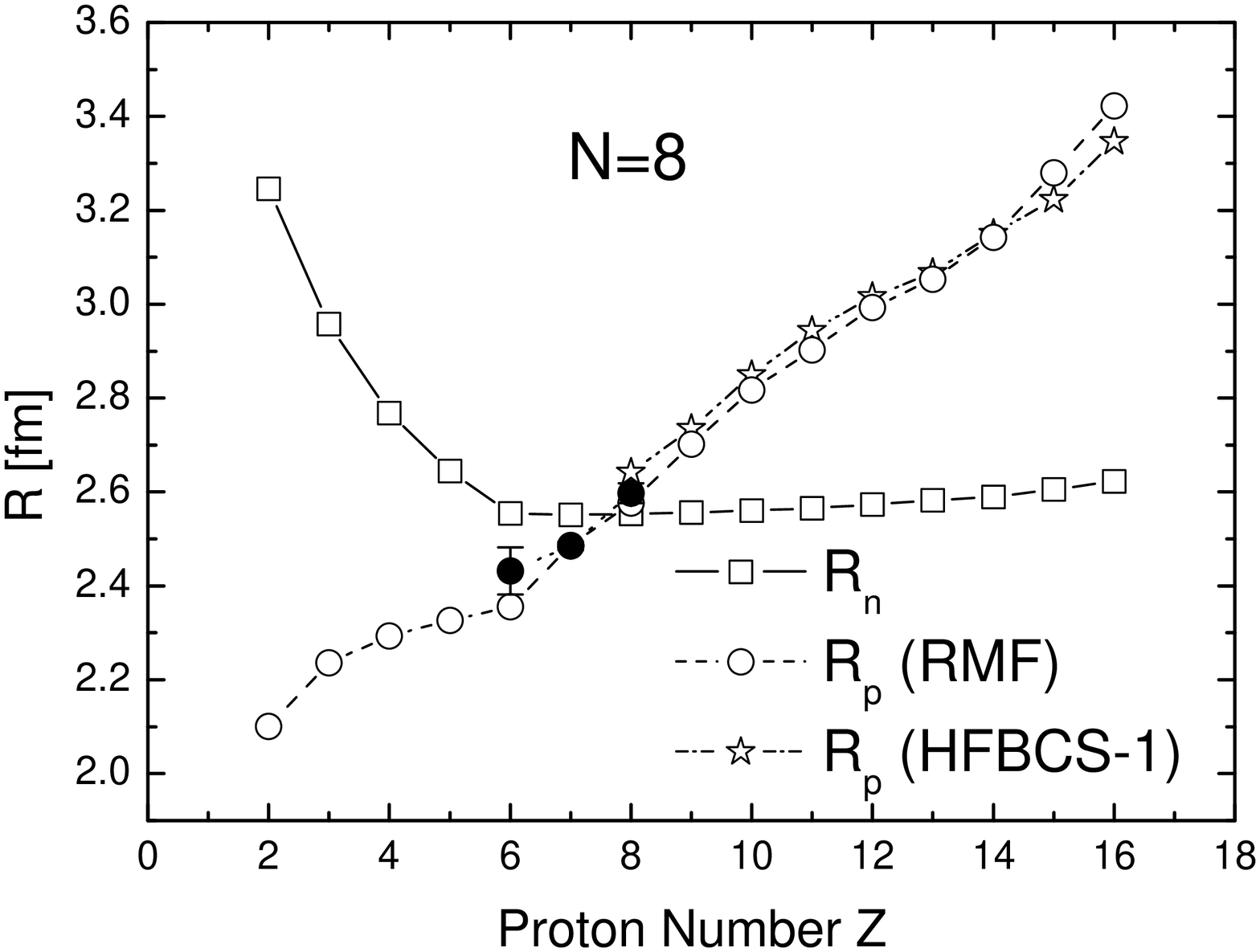}%
\includegraphics[scale=0.23]{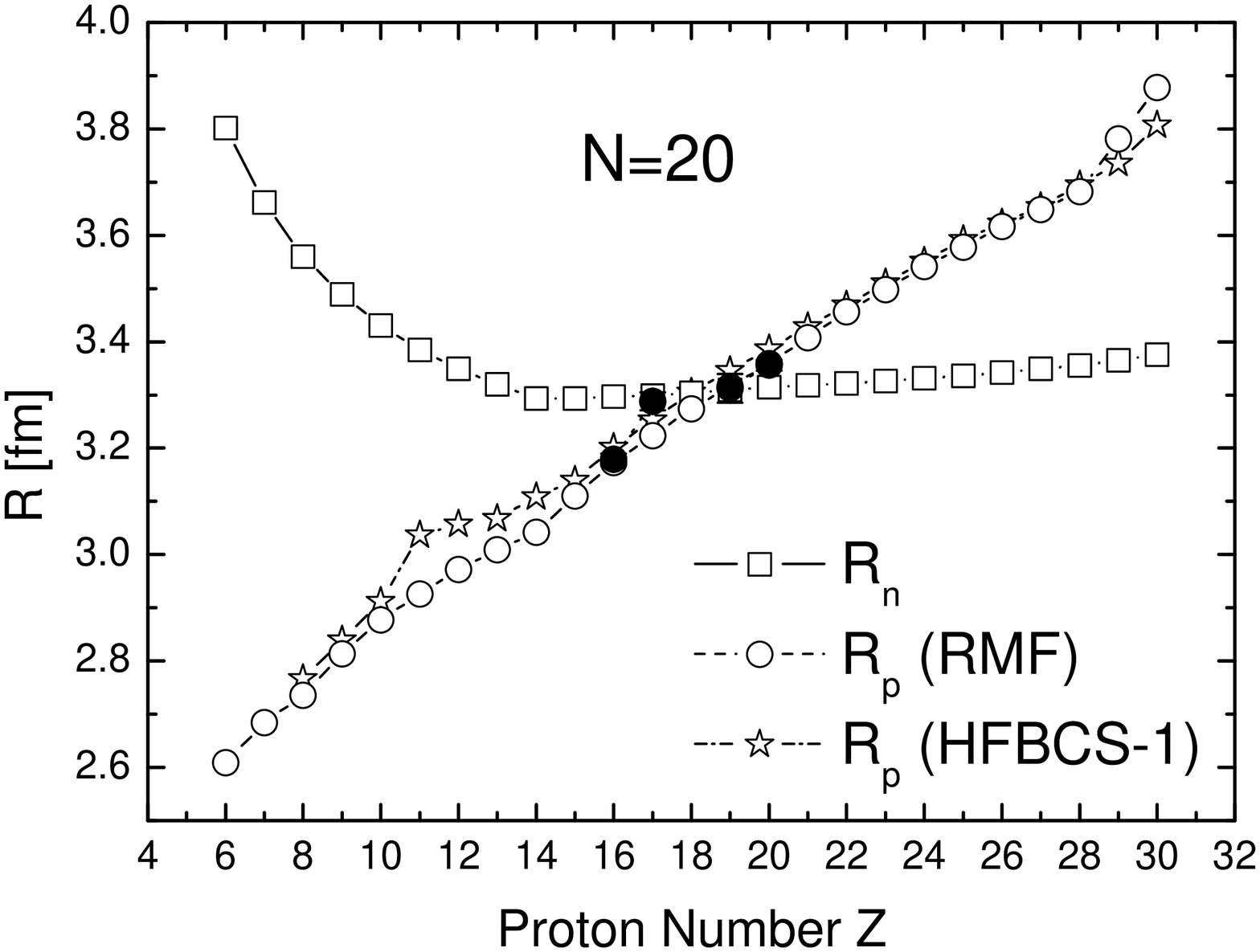}%
\includegraphics[scale=0.23]{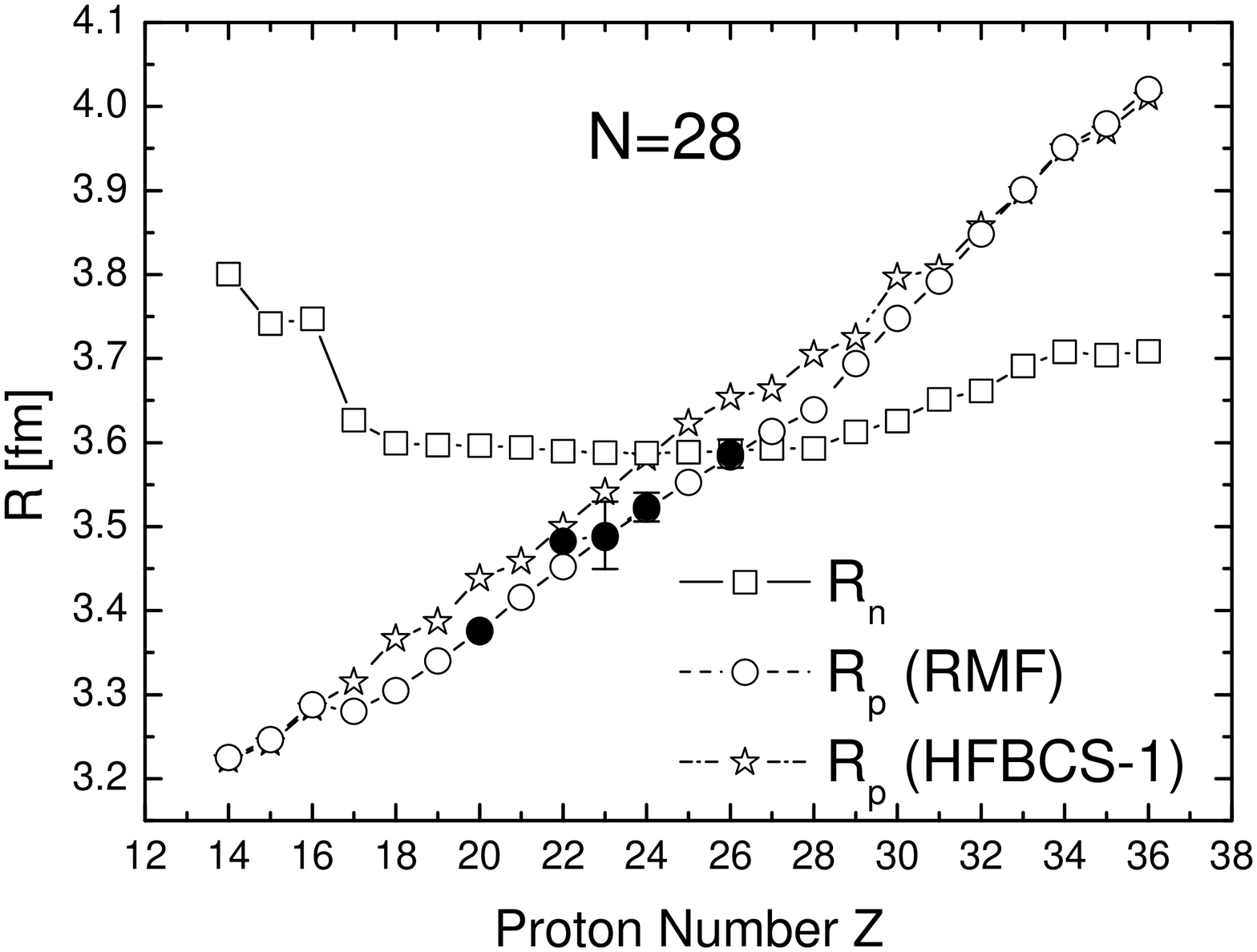}
\includegraphics[scale=0.23]{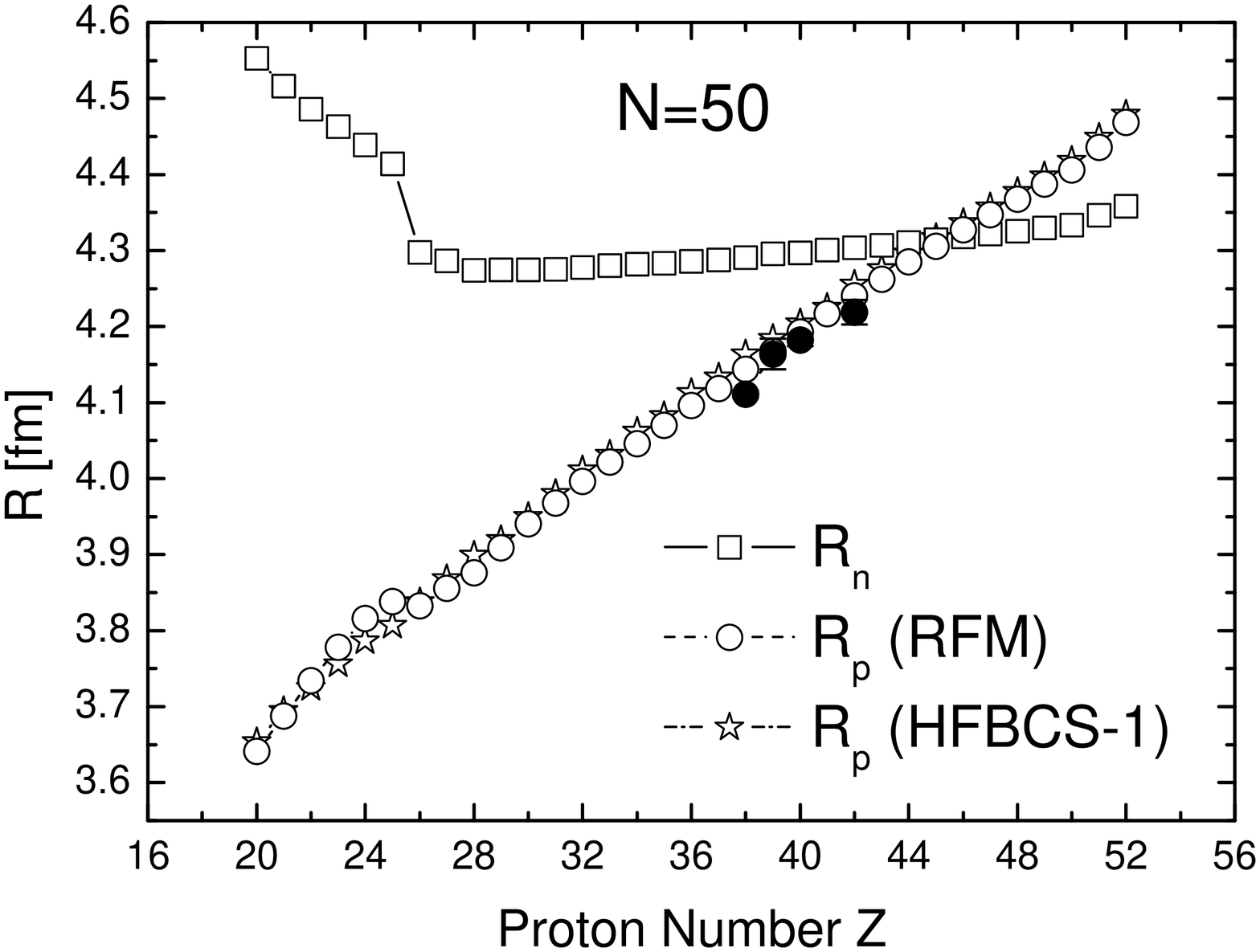}%
\includegraphics[scale=0.23]{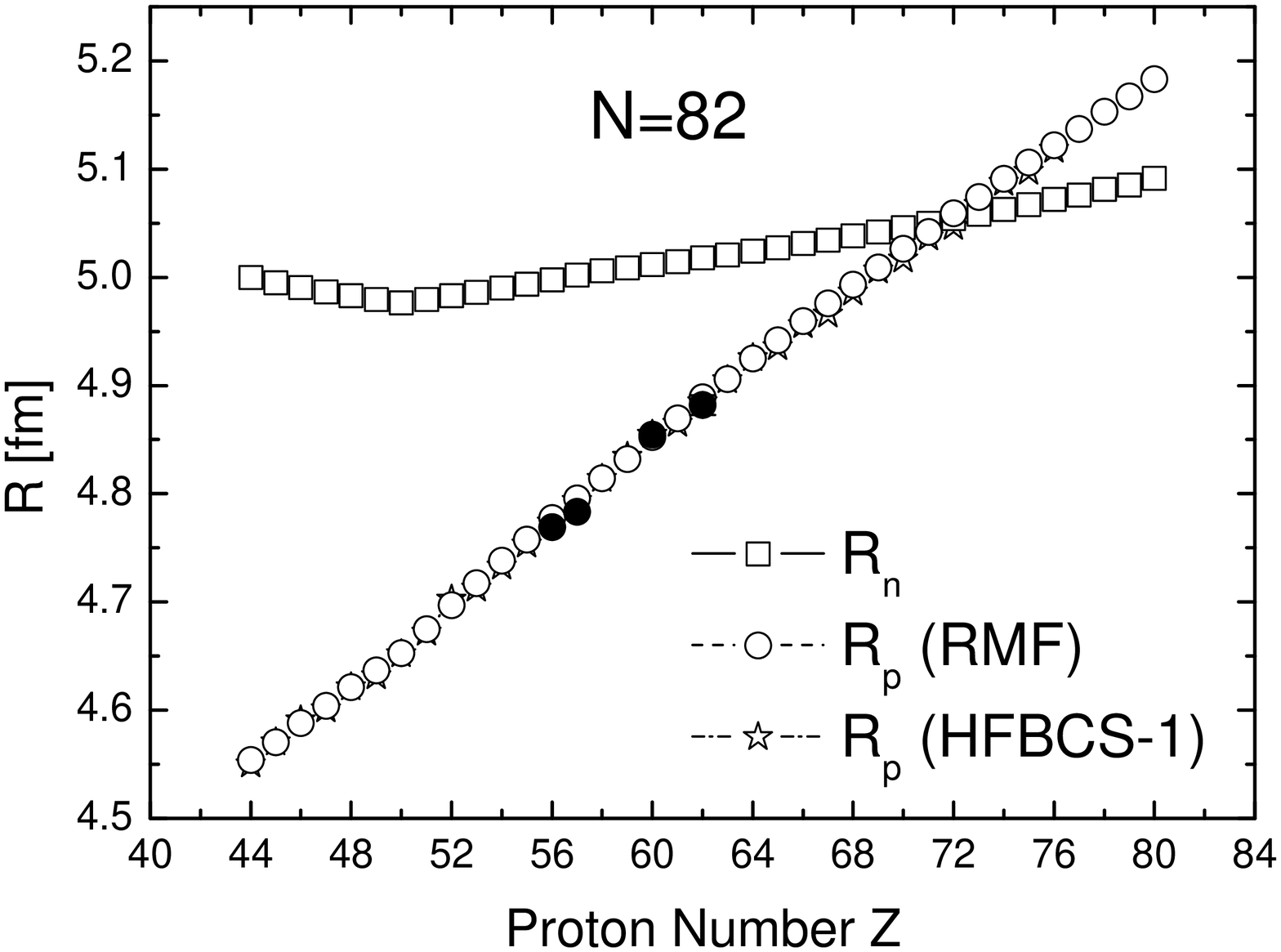}%
\includegraphics[scale=0.23]{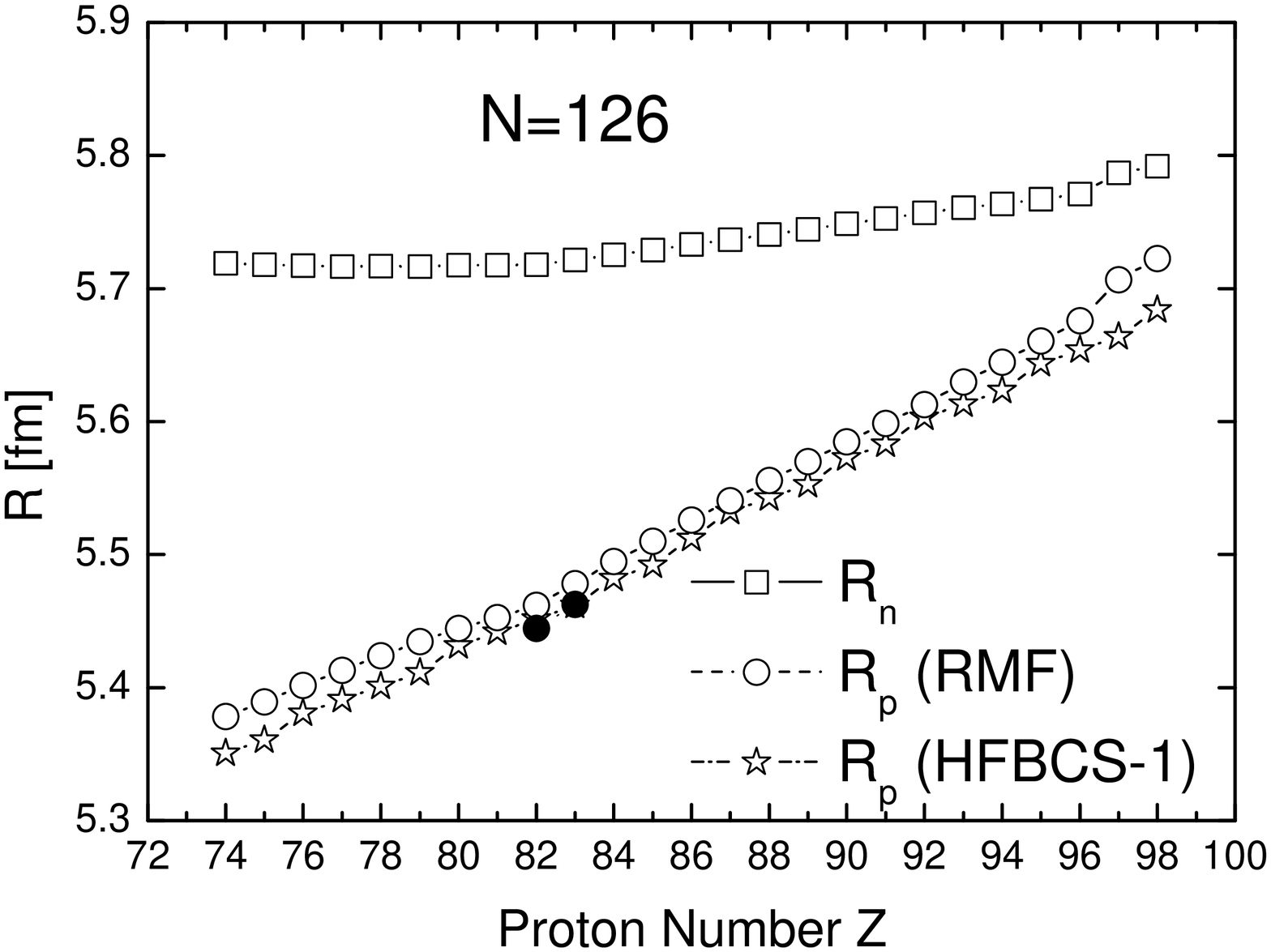}
\caption{The rms neutron and proton radii, $R_n$ and $R_p$, of the
$N=8$, 20, 28, 50, 82, and 126 isotones as functions of the proton
number, $Z$. The results obtained from the deformed RMF+BCS
calculations with the TMA parameter set (open circle) are compared
with the available experimental data (solid circle with error bar)
\cite{vries.87} and those of the HFBCS-1 mass formula (open star)
\cite{hfbcs1.01}.}
\end{figure*}
\begin{figure*}[htpb!]
\centering
\includegraphics[scale=0.23]{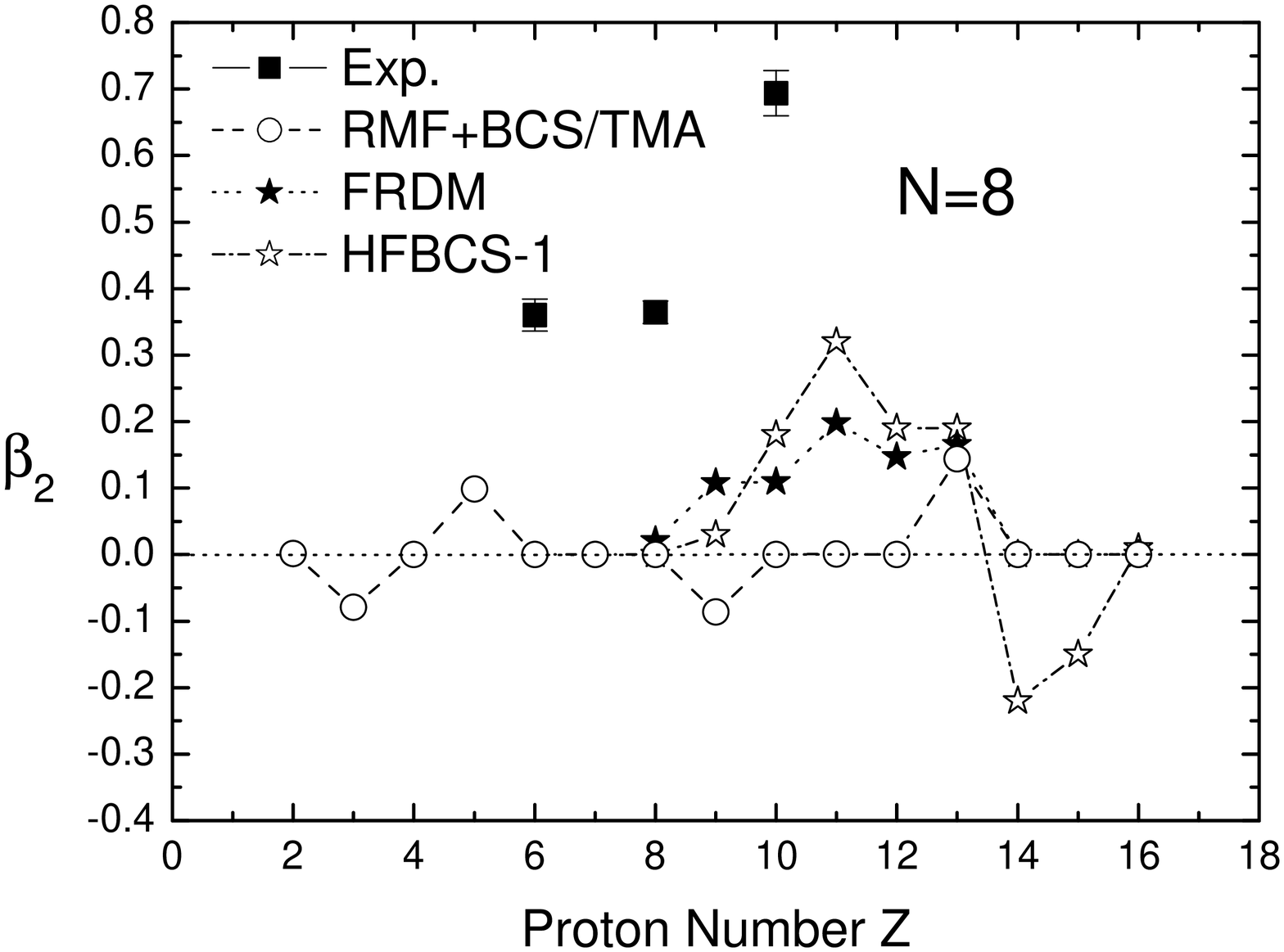}%
\includegraphics[scale=0.23]{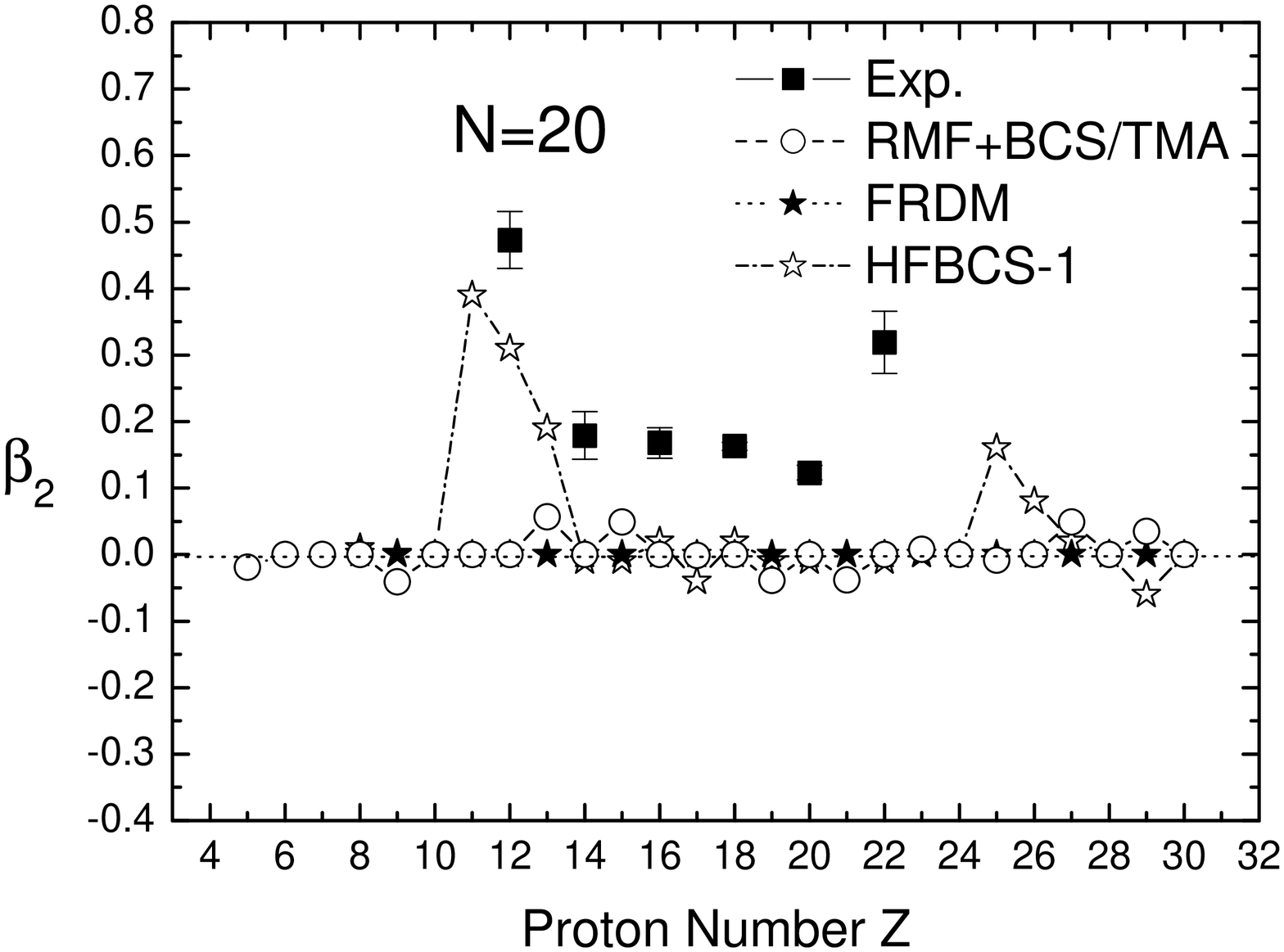}%
\includegraphics[scale=0.23]{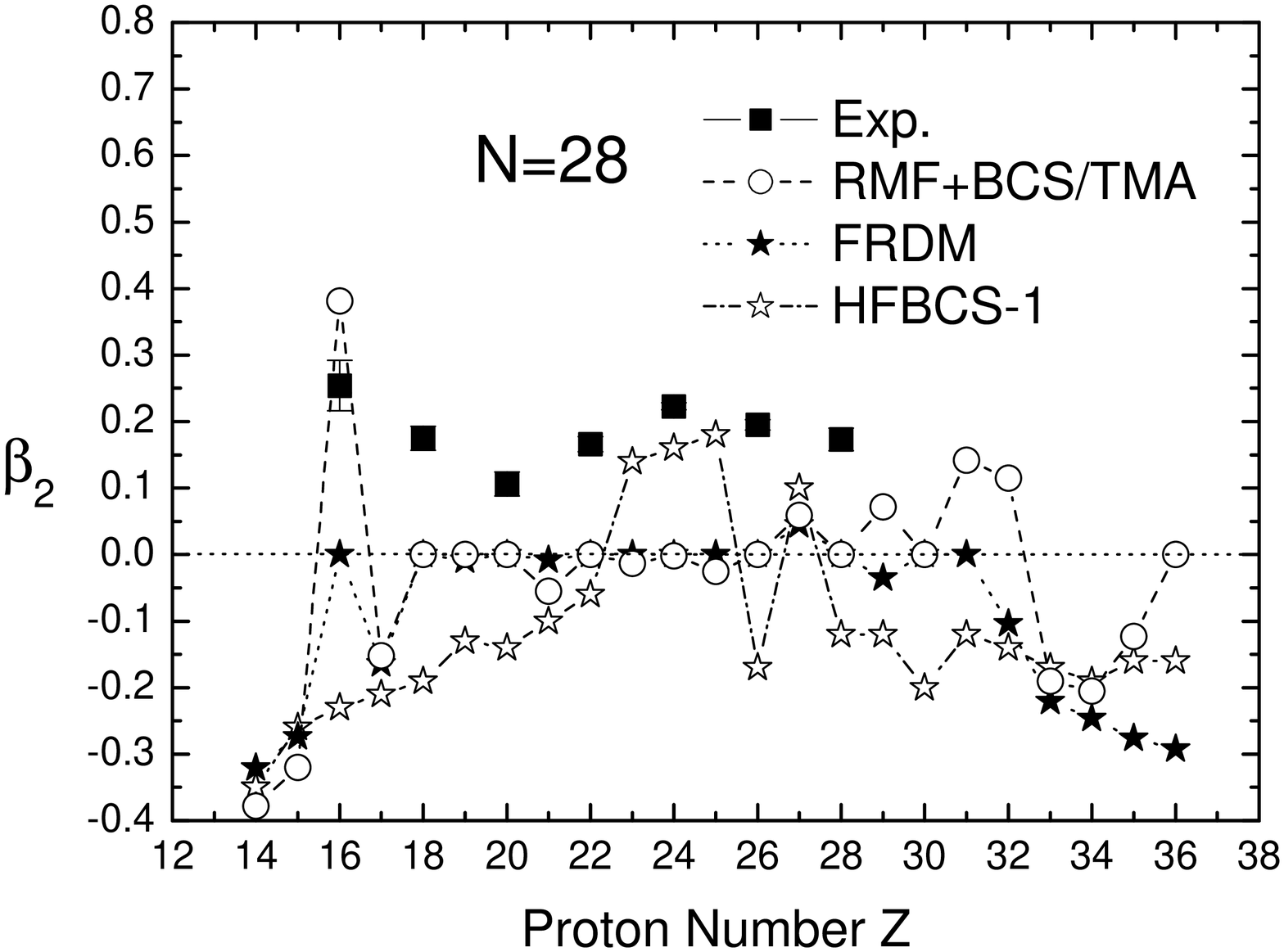}
\includegraphics[scale=0.23]{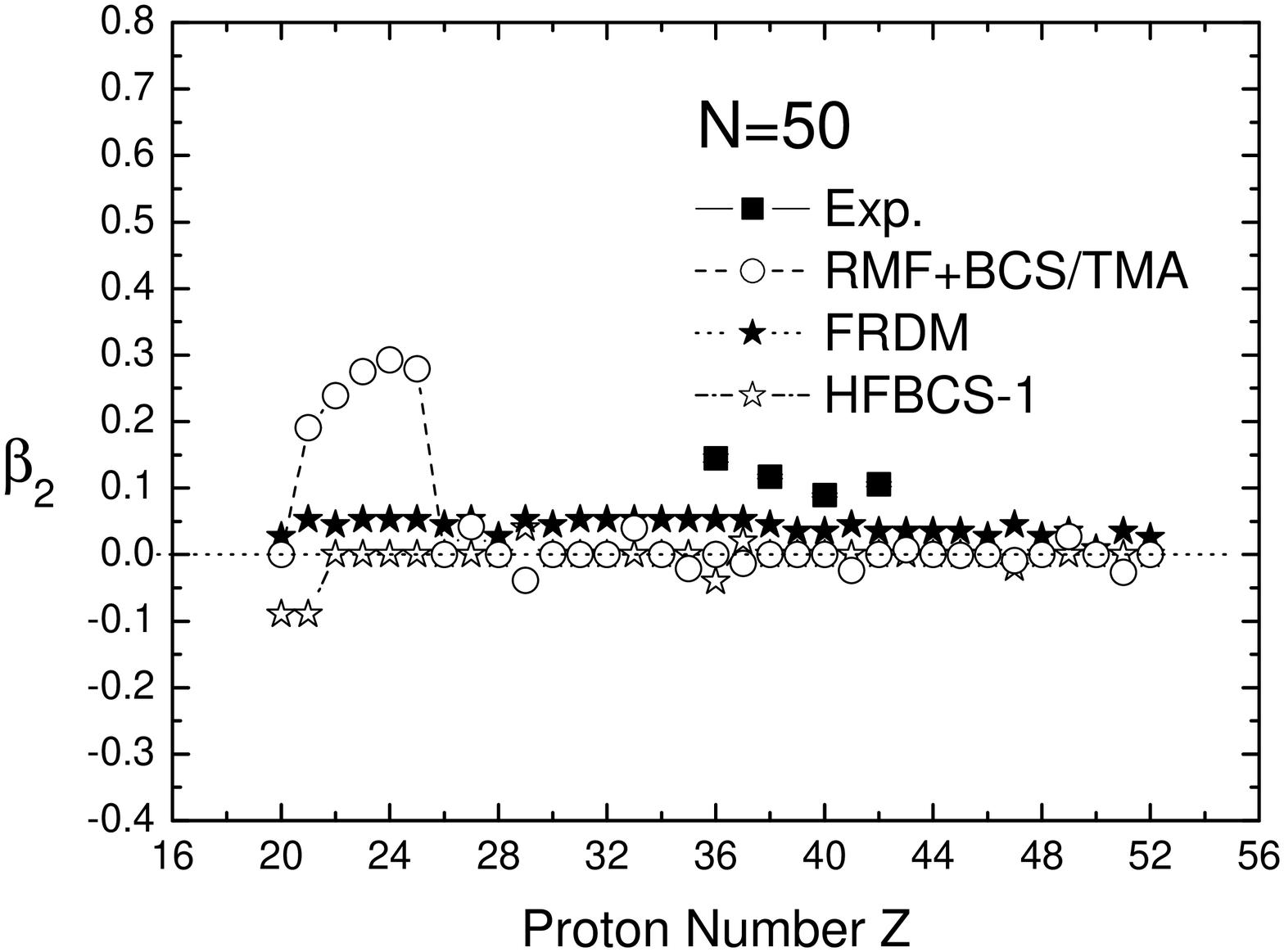}%
\includegraphics[scale=0.23]{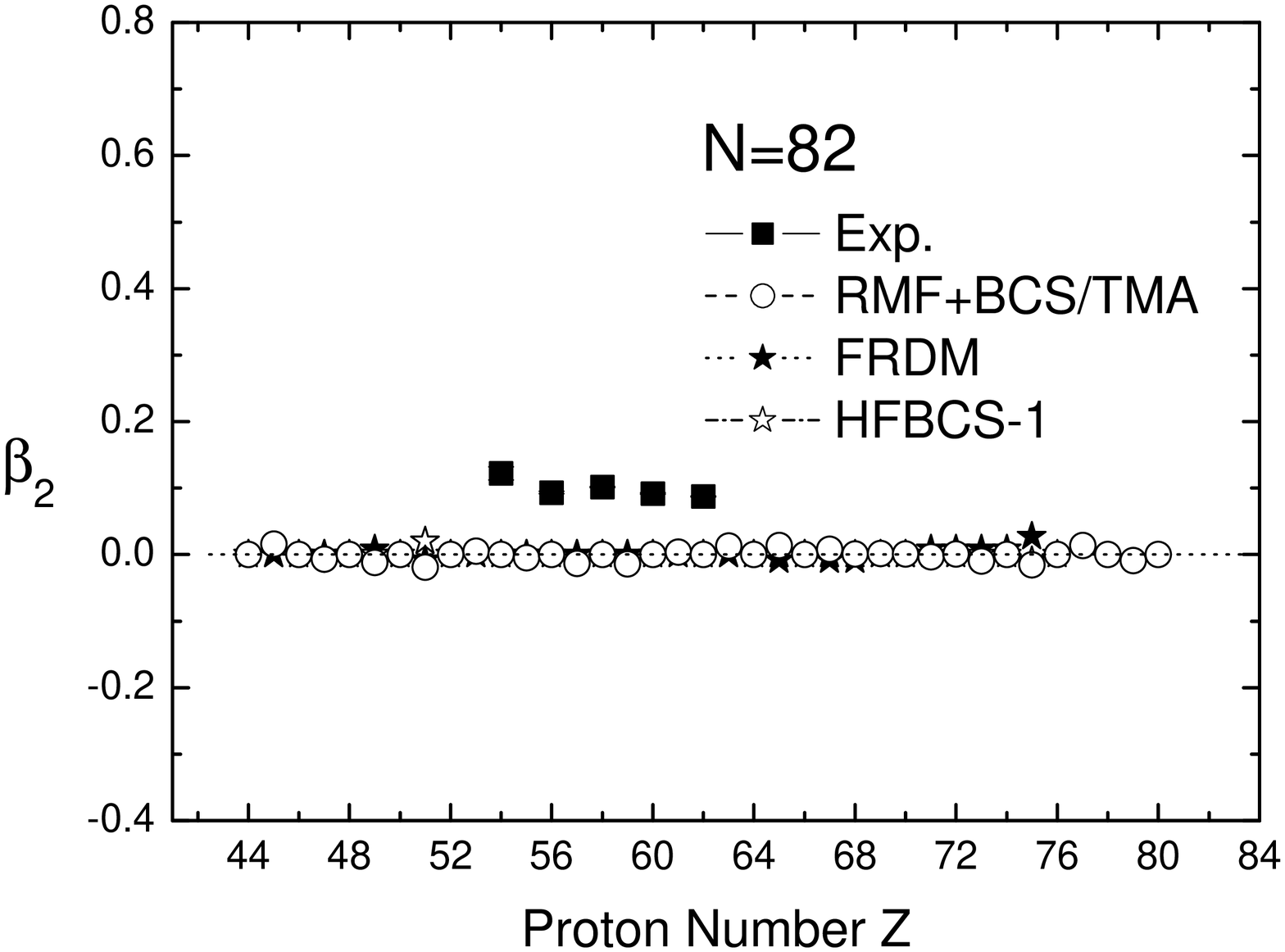}%
\includegraphics[scale=0.23]{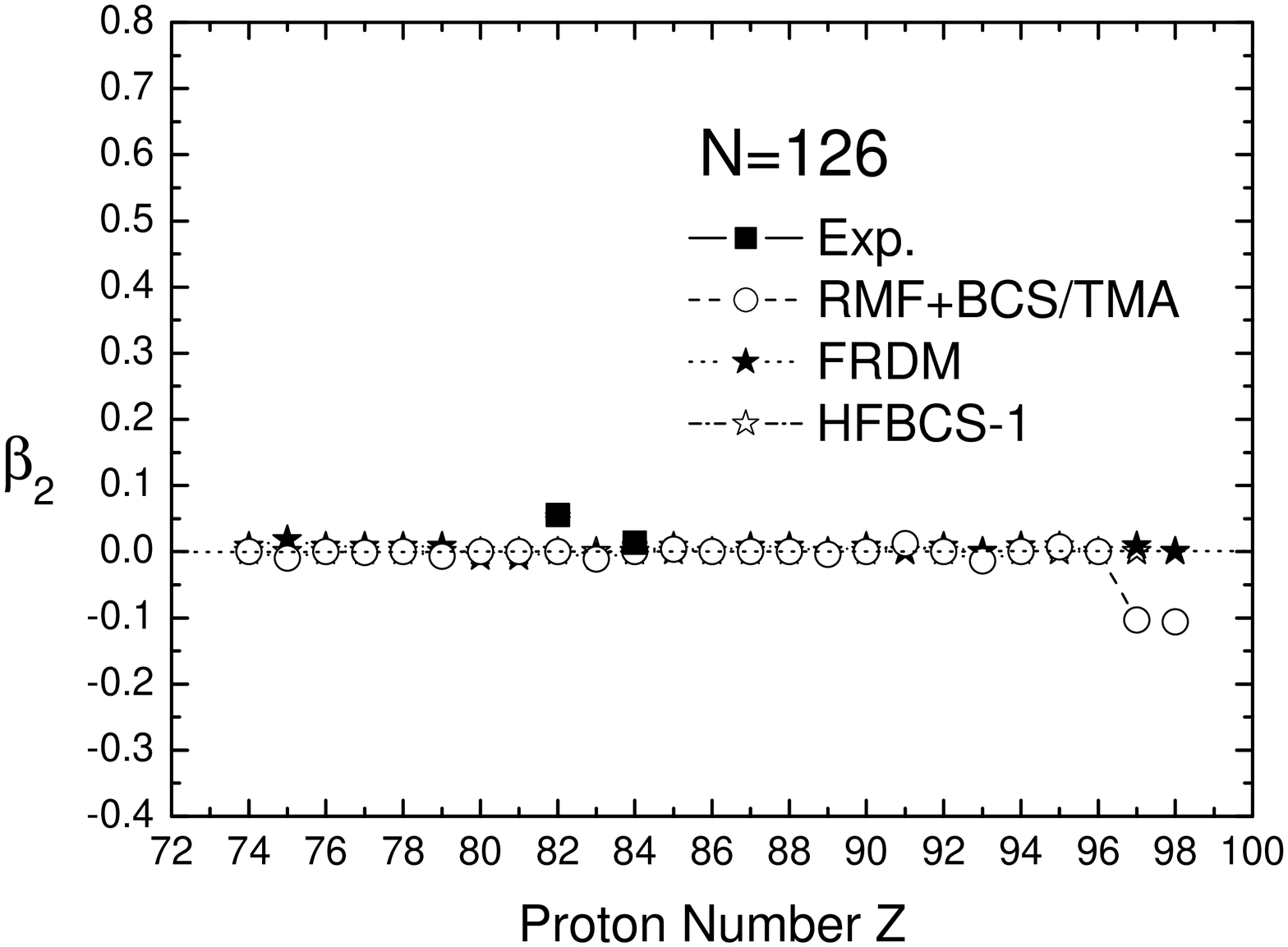}
\caption{The mass quadrupole deformation parameters, $\beta_2$, of
the $N=8$, 20, 28, 50, 82, and 126 isotones as functions of the
proton number, $Z$. The results obtained from the deformed RMF+BCS
calculations with the TMA parameter set (open circle) are compared
with the predictions of the FRDM model (solid star)
\cite{moller.95}, those of the HFBCS-1 mass formula (open star)
\cite{hfbcs1.01}, and the available experimental data (solid
square with error bar) extracted from the $B(E2:0^+\rightarrow
2^+)$ values \cite{raman.01}. Note that the extracted deformation
parameter does not mean that the corresponding nucleus is really
deformed.}
\end{figure*}

\section{Nuclear radii}

The difference between neutron and proton root-mean-square (rms)
radii,
\begin{equation}
R_{n(p)}=\langle r^2_{n(p)}\rangle^{1/2}= \left[\frac{\int
\rho_{n(p)} r^2d{\bf r}}{\int \rho_{n(p)} d{\bf r}}\right]^{1/2},
\end{equation}
where $\rho_{n(p)}$ is the neutron (proton) density distribution,
can be very large for nuclei with exotic isospin ratios, thus
forms the so-called neutron (proton) skin \cite{geng.032}, and
even neutron (proton) halo \cite{sand.03,meng.98,meng98prl}. The
neutron rms radii (empty square) and proton rms radii (empty
circle) of the $N = 8$, 20, 28, 50, 82 and 126 isotones are
plotted in Fig. 3, where the experimental proton radii (solid
circle with error bar) are extracted from the experimental charge
 radii \cite{vries.87} by the formula,
\begin{equation}
R^2_p=R^2_c-0.64\mbox{ fm}^2.
\end{equation}
Since the HFBCS-1 mass formula has achieved an unprecedented
success in describing nuclear charge radii \cite{buchinger.01}
compared with the compilation of Nadjakov et al.
\cite{nadjakov.94}, its results are also shown for comparison
(open stars).

 In Fig. 3, it is clearly seen that the agreement between our
calculations and the experimental data is remarkably good. For
$^{37}_{17}$Cl$_{20}$ and $^{50}_{22}$Ti$_{28}$, the experimental
data are somewhat larger than our theoretical predictions. Also,
the proton radii of two nuclei in the $N=126$ isotonic chain,
$^{208}_{82}$Pb$_{126}$ and $^{209}_{83}$Bi$_{126}$, seem to be
overestimated a bit in our calculations. Some kinks in Fig. 3 are
due to the sudden changes of deformations of the corresponding
nuclei (see Fig. 4), including the one at $Z=16$ in the $N=28$
isotonic chain, the one at $Z=25$ in the $N=50$ isotonic chain,
and the one at $Z=97$ in the $N=126$ isotonic chain. It should be
noted that the last few nuclei in each isotonic chain shown in
Fig. 3 are unbound in our calculations with separation energies
$-5<S_{p(2p)}<0$ MeV (see Fig. 2), thus the sudden increases of
the proton radii do not necessarily imply appearances of proton
halos.

The agreement between our results and those of the HFBCS-1
\cite{hfbcs1.01} mass formula is also very good in general,
particularly for those nuclei in the $N=50$ and $82$ isotonic
chains. The discrepancies for some nuclei with $Z\approx12$ in the
$N=20$ isotonic chain are due to the deformation effects (see Fig.
4). And the same is true for the nuclei with $17\le Z\le 30$ in
the $N=28$ isotonic chain, but the experimental data seems to be
in favor of our results. For all the nuclei in the $N=126$
isotonic chain, the results of the HFBCS-1 mass formula are a bit
smaller than our results and therefore are closer to the
experimental data.

One more thing we note is that unlike proton radii, which go up
almost monotonously with increasing proton number, neutron radii
usually go down first, then go up slowly. This phenomenon is more
obvious in the $N=8$, 20, 28, and 50 isotonic chains. The
underlying reason is not difficult to understand. For
proton-deficient nuclei with small $Z$, the protons occupy
``inner'' single-particle states with much smaller rms radius than
the neutron rms radius. Therefore, an extra proton added to these
``inner'' single-particle states attracts the neutrons closer to
the nuclear center. Consequently, the neutron rms radius decreases
until Pauli principle prohibits further protons from occupying
these ``inner'' single-particle states. With more protons added,
the opposite becomes true and the neutron rms radius begins to
increase, though very slowly.

\section{Deformations}

Deformation is another important property of nuclei. It also could
be used as one of the indicators of magicity conserving or losing
of the corresponding nucleus. In our previous work
\cite{geng.033}, we found that Sn isotopes are deformed in the
neutron-rich region, which could be viewed as the magicity losing
of proton magic number $Z=50$. However, experimentally, it is
difficult to obtain deformation knowledge of nuclei directly. One
of the most usual methods is to extract deformation parameters
$\beta_2$ from the $B(E2:0^+\rightarrow 2^+)$ values by using the
assumption of deformation \cite{raman.01}. Hence, all the nuclei
have a finite deformation and also the analysis is limited to
even-even nuclei. Due to this fact, we have to treat the extracted
deformation parameters with care. For most cases, if the extracted
deformation parameter is small, the nucleus is expected to be
spherical.

In Fig. 4, we plot the mass deformation parameters, $\beta_2$, for
all the $N=8$, 20, 28, 50, 82, and $126$ isotones as functions of
the proton number $Z$. The experimental data are extracted from
the $B(E2:0^+\rightarrow 2^+)$ values given in Ref.
\cite{raman.01}. The predictions of the FRDM mass formula
\cite{moller.95} and those of the HFBCS-1 mass formula
\cite{hfbcs1.01} are also shown for comparison. From Fig. 4, it is
easily seen that the agreement between the results of our
calculations and those of the other two methods is remarkably
good. All these three methods obtain essentially the same results
for all the $N=50$, 82 and 126 isotones. For the $N=8$ isotonic
chain, FRDM predicts $^{18}_{10}$Ne$_8$, $^{19}_{11}$Na$_8$, and
$^{20}_{12}$Mg$_8$ to be slightly deformed, while in our
calculations, these nuclei are spherical. On the contrary, the
HFBCS-1 mass formula predicts all the nuclei with $Z\ge 10$ to be
slightly deformed except for $^{24}_{16}$S$_8$. For the $N=20$
isotonic chain, both our calculations and the FRDM method show
that all the nuclei are spherical, while the HFBCS-1 mass formula
predicts the nuclei with $Z=11,12,13$, and 25 are deformed, which
seems to be closer to the experimental data. For the $N=28$
isotonic chain, our calculations predict $^{44}_{16}$S$_{28}$ to
be deformed, which is consistent with both the experimental data
and the predictions of the RHB method \cite{lala.99}, while FRDM
predicts it to be spherical. These two methods also differ in the
predictions for the proton drip-line nuclei,
$^{63}_{35}$Br$_{28}$, $^{64}_{36}$Kr$_{28}$,
$^{213}_{97}$Bk$_{126}$, and $^{214}_{98}$Kr$_{126}$. We should
note that these nuclei are already unstable against proton
emission in our calculations (see Fig. 2). On the contrary, the
HFBCS-1 method agrees with our calculations at both ends of this
isotonic chain, but differs a bit in the middle: It predicts all
the nuclei with $17\le Z \le 30$ to be slightly deformed. That
could be the reason why there are some discrepancies between our
results and those of the HFBCS-1 mass formula, as shown in Fig. 3.
For the $N=50$ isotonic chain, the main differences between the
predictions of these three methods exist in the region with
$20<Z<26$. Our calculations show that these nuclei are deformed,
while both FRDM and HFBCS-1 predict them to be nearly spherical.

Except for nuclei with $Z<18$ and $Z>30$ in the $N=28$ isotonic
chain, and nuclei with $20<Z<26$ in the $N=50$ isotonic chain, all
the other nuclei in our calculations are more or less spherical.
Although, at first sight, the agreement between our calculations
(including the predictions of the FRDM model and the HFBCS-1 mass
formula) and the experimental data is not very good, we should
keep in mind the limitations of the extracting method that we
obtain the experimental deformation parameters. Within such a
method, even the well-known spherical nuclei, $^{16}_8$O$_8$ and
$^{40}_{20}$Ca$_{20}$, are predicted to be deformed
\cite{raman.01} (see Fig. 4). Therefore, the experimental data
should be compared with caution. However, combined with other
experimental knowledge, such as the reduced electric transition
probability, $B(E2)$, and $2_1^+$ energy, we can decide whether a
nucleus is really deformed. The corresponding quantities for
$^{18}_{10}$Ne$_8$, $^{32}_{12}$Mg$_{20}$, and
$^{42}_{22}$Ti$_{20}$ are 0.0269(26) $e^2b^2$, 1887.3(2) KeV;
0.039(7) $e^2b^2$, 885.5(7) KeV; 0.087(25) $e^2b^2$, 1554.9(8)
KeV; respectively, which indicate these nuclei are really deformed
\cite{raman.01}. The fact that our calculations did not find
deformed ground-state configurations for these nuclei is probably
due to the broken rotational symmetry in the relativistic mean
field theory, which will be discussed below .

One of the long unsolved problems in the relativistic mean field
theory is that it cannot obtain the deformed ground-state
configurations for $^{32}$Mg, which is known to be strongly
deformed experimentally \cite{raman.01}, and other $N=20$ isotones
around $Z=12$, including $^{31}$Na and $^{30}$Ne
\cite{geng.032,lala.01}. On the other hand, the AMD method
\cite{horiuchi.02}, the shell-model calculations
\cite{caurier.98,utsuno.99}, and the angular momentum projected
generator coordinate method \cite{guzman.00} can obtain a deformed
ground-state configuration for $^{32}$Mg. In the shell-model
calculations \cite{caurier.98,utsuno.99}, it is demonstrated that
the $2p-2h$ intruders dominate the ground states of $^{30}$Ne,
$^{31}$Na, and $^{32}$Mg, therefore lead to deformed ground states
for these nuclei. While the studies of the AMD method and the
angular momentum projected generator coordinate method clearly
show that the zero-point energy associated with the restoration of
the broken rotational symmetry \cite{reinhard.99} is indispensable
for a correct description of the ground state of $^{32}$Mg.

As mentioned above, we have performed the quadrupole constrained
calculations for every nucleus in our present work in order to
obtain its potential energy surface and determine the
corresponding ground state. In Fig. 5, we plot the potential
energy surface of $^{32}$Mg as a function of the deformation
parameter $\beta_2$. For comparison, the results calculated with
the parameter sets NL3 and NL-Z2 are also shown. For all the three
parameter sets, an indication of a second minimum around
$\beta_2\approx0.5$ is clearly seen (though very ``soft'' for TMA
and NL3), which agrees well with the unprojected AMD
\cite{horiuchi.02} and HFB \cite{caurier.98} calculations. (Here
it should be noted that the total energy calculated with NL-Z2
seems to deviate from those obtained with TMA and NL3 somewhat,
therefore does not agree with experiment very well. It could be
due to the fact that NL-Z2 is aimed at heavy and superheavy nuclei
but not light nuclei like $^{32}$Mg. Also, we should note that TMA
and NL3 adopt the same scheme of center-of-mass correction, while
NL-Z2 uses the so-called microscopic center-of-mass correction
\cite{bender.99}.) Therefore, we conclude that to obtain the
deformed ground-state configuration for $^{32}$Mg, including
$^{18}$Ne and $^{44}$Ti, the broken rotational symmetry in the
relativistic mean field theory has to be restored. Since so far,
to our knowledge, such a calculation is still missing in the RMF
method, no wonder no RMF calculations have succeeded in obtaining
deformed ground-state configurations for these nuclei. We also
consider this as one motivation to introduce the angular momentum
projection method into the present RMF model to restore the broken
rotational symmetry as one of our next works.

In Ref. \cite{dillmann.03}, the authors argued that their
experimental high $Q_\beta$ value, $8344^{+165}_{-157}$ KeV, for
$^{130}$Cd is a direct signature of an $N=82$ shell quenching
below $^{132}$Sn. While in our calculations, from the point of
view of deformations, there is no indications of shell quenching
for the whole $N=82$ isotonic chain, even in the neutron-rich
side. In our calculations, the ground-state binding energy of
$^{130}$In, $E=1080.966$ MeV ($\beta_2=0.041$), together with the
binding energy of $^{130}$Cd, $E=1072.897$ MeV ($\beta_2=0.0$),
give a $Q_\beta$ of 8849 KeV, which falls into the category of
``quenched models'' defined in Ref. \cite{dillmann.03}. Therefore,
whether or not the $N=82$ shell is quenched in the neutron-rich
region needs a more careful study both experimentally and
theoretically.

\begin{figure}[t]
\centering
\includegraphics[scale=0.35]{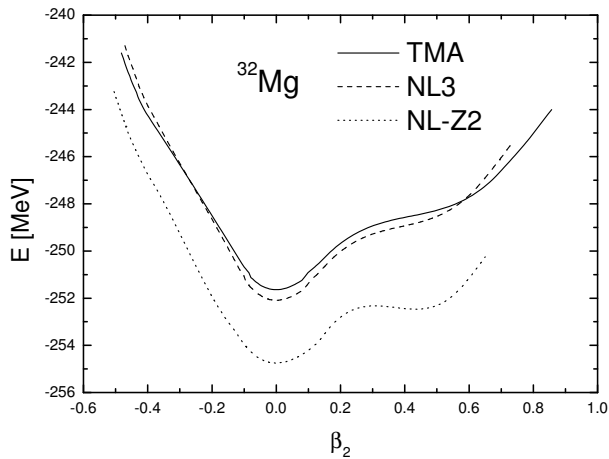}%
\caption{The potential energy surfaces of $^{32}$Mg as functions
of the deformation parameter, $\beta_2$. The solid, dashed, and
dotted lines correspond to the results calculated with the
parameter sets TMA, NL3, and NL-Z2, respectively.}
\end{figure}
\section{Conclusions}

Nuclei with magic numbers are very important in the study of
nuclear physics, both theoretically and experimentally. The
localized feature of ``magic numbers'' has received more and more
attention due to its importance to astrophysical problems and
recent experimental developments to produce exotic nuclei in the
laboratory. In the present work, we have adopted the relativistic
mean field theory to study neutron magic nuclei with six classical
magic numbers $N=8$, 20, 28, 50, 82, and $126$. The relativistic
mean field theory that has proved to be very successful in
describing many nuclear properties, due to its natural spin-orbit
description and only a few parameters, is considered to be an
ideal model to study not only stable nuclei but also exotic
nuclei. In order to study all those nuclei in a proper way, we
must take into account pairing correlations and deformation
effects simultaneously. For nuclei with odd number of nucleons,
the blocking effect must also be treated properly
\cite{geng.prcr}. All these requirements make the calculation very
complicated and time consuming, so up to now a systematic study in
the relativistic mean field theory including all these neutron
magic numbers is still missing. However, due to the importance of
these nuclei to various physical problems, such a study is in
urgent need. This work is also part of our series of efforts to
study the magicity of magic numbers in the relativistic mean field
theory.

In our systematic study of the neutron magic isotonic chains, we
have found a new proton magic number $Z=6$ in the $N=8$ isotonic
chain. The shell closure feature at $Z=14$ in the $N=20$ isotonic
chain is also reproduced very well. We have failed, however, to
obtain deformed ground-state configurations for nuclei around
$Z=12$ in the $N=20$ isotonic chain, which have been attributed to
the broken rotational symmetry. $N=28$ has been predicted to lose
its magicity in the proton-deficient side of the isotonic chain,
i.e. they are found to be deformed. $N=50$ seems to conserve its
magicity in the proton-rich side while lose its magicity in the
proton-deficient side. This is in agreement with its proton
counterpart, $Z=50$, which is also found to lose its magicity in
the neutron-rich side. For the $N=82$ isotonic chain, we have
confirmed that there is a (sub)shell closure at $Z=58$, but the
different deviation trends in the $Z>58$ and $Z<58$ regions seem
to imply that some features are missed and/or mistreated in the
relativistic mean field theory. For the $N=126$ isotonic chain,
both the two- and one-proton separation energies are a bit
overestimated. Our calculations also indicate a new (sub)shell
closure at $Z=92$. The binding energies agree well with experiment
around $^{208}$Pb and become larger than experiment with
increasing proton number $Z$. The use of two other parameter sets,
NL3 and NL-Z2, does not change the above conclusion. The predicted
two-proton and one-proton drip-line nuclei for the $N=8$, 20, 28,
50, 82, and 126 isotonic chains are $^{20}_{12}$Mg$_{8}$;
$^{46}_{26}$Fe$_{20}$, $^{44}_{24}$Cr$_{20}$;
$^{58}_{30}$Zn$_{28}$, $^{56}_{28}$Ni$_{28}$;
$^{100}_{50}$Sn$_{50}$; $^{156}_{74}$W$_{82}$,
$^{152}_{70}$Yb$_{82}$; $^{220}_{94}$Pu$_{126}$;
$^{218}_{92}$U$_{126}$; respectively.

This work was partly supported by the Major State Basic Research
Development Program Under Contract Number G2000077407 and the
National Natural Science Foundation of China under Grant No.
10025522, 10221003 and 10047001.


\end{document}